\RequirePackage{lineno}
\documentclass[aps,prl,10pt,twocolumn,superscriptaddress,longbibliography,nofootinbib]{revtex4-2}

\usepackage{graphicx}
\usepackage{amssymb,amsmath,amsfonts,gensymb}
\usepackage{bm,hyperref}
\usepackage{color}
\usepackage{ulem}
\usepackage{bbm}
\usepackage{subfigure}
\usepackage{dcolumn}
\usepackage{mathtools}
\usepackage{pifont}
\usepackage[symbol]{footmisc}
\usepackage{mhchem}

\newcommand{\ket}[1]{\vert #1 \rangle}

\newcommand{\eV}{{\text{eV}}}
\newcommand{\eVA}{{\text{eV}\cdot \text{\AA}}}
\newcommand{\bk}{\mathbf{k}}
\newcommand{\bq}{\mathbf{q}}

\newcommand{\bd}{\mathbf{d}}
\newcommand{\btk}{\widetilde{\mathbf{k}}}
\newcommand{\btq}{\widetilde{\mathbf{q}}}
\newcommand{\br}{\mathbf{r}}

\newcommand{\dop}{\hat{d}}

\def\Blue#1{\textcolor{blue}{#1}}
\definecolor{BLUE}{rgb}{0,0,1}
\def\BLUE#1{\textcolor{BLUE}{#1}}
\def\I{\uppercase\expandafter{\romannumeral 1}}
\def\II{\uppercase\expandafter{\romannumeral 2}}
\def\III{{\uppercase\expandafter{\romannumeral 3}}}
\def\IV{{\uppercase\expandafter{\romannumeral 4}}}
\def\V{{\uppercase\expandafter{\romannumeral 5}}}
\def\VI{{\uppercase\expandafter{\romannumeral 6}}}
\def\VII{{\uppercase\expandafter{\romannumeral 7}}}
\def\i{\lowercase\expandafter{\romannumeral 1}}
\def\ii{\lowercase\expandafter{\romannumeral 2}}
\def\iii{{\lowercase\expandafter{\romannumeral 3}}}

\def\a{\mathbf{a}}
\def\b{\mathbf{b}}
\def\p{\mathbf{p}}

\def\k{\mathbf{k}}

\def\G{\mathbf{G}}
\def\Q{\mathbf{Q}}
\def\br{\mathbf{r}}
\def\bd{\mathbf{d}}
\def\kt{\widetilde{\mathbf{k}}}
\def\q{\mathbf{q}}

\def\qt{\widetilde{\mathbf{q}}}
\def\nn{\nonumber \\}

\def\R{\mathbf{R}}

\makeatletter
\def\@ssect@ltx#1#2#3#4#5#6[#7]#8{%
	\def\H@svsec{\phantomsection}%
	\@tempskipa #5\relax
	\@ifdim{\@tempskipa>\z@}{%
		\begingroup
		\interlinepenalty \@M
		#6{%
			\@ifundefined{@hangfroms@#1}{\@hang@froms}{\csname @hangfroms@#1\endcsname}%
			{\hskip#3\relax\H@svsec}{#8}%
		}%
		\@@par
		\endgroup
		\@ifundefined{#1smark}{\@gobble}{\csname #1smark\endcsname}{#7}%
	}{%
		\def\@svsechd{%
			#6{%
				\@ifundefined{@runin@tos@#1}{\@runin@tos}{\csname @runin@tos@#1\endcsname}%
				{\hskip#3\relax\H@svsec}{#8}%
			}%
			\@ifundefined{#1smark}{\@gobble}{\csname #1smark\endcsname}{#7}%
			\addcontentsline{toc}{#1}{\protect\numberline{}#8}%
		}%
	}%
	\@xsect{#5}%
}%
\makeatother

\begin{document}

%\linenumbers
	
\title{Correlated states in charge-transfer heterostructures based on rhombohedral multilayer graphene}

\author{Yanran Shi}
\affiliation{School of Physical Science and Technology, ShanghaiTech Laboratory for Topological Physics, State Key Laboratory of Quantum Functional Materials, ShanghaiTech University, Shanghai 201210, China}

\author{Min Li}
\affiliation{School of Physical Science and Technology, ShanghaiTech Laboratory for Topological Physics,  State Key Laboratory of Quantum Functional Materials, ShanghaiTech University, Shanghai 201210, China}

\author{Xin Lu}
\email{lvxin@shanghaitech.edu.cn}
\affiliation{School of Physical Science and Technology, ShanghaiTech Laboratory for Topological Physics, State Key Laboratory of Quantum Functional Materials, ShanghaiTech University, Shanghai 201210, China}

\author{Jianpeng Liu}
\email{liujp@shanghaitech.edu.cn}
\affiliation{School of Physical Science and Technology, ShanghaiTech Laboratory for Topological Physics, State Key Laboratory of Quantum Functional Materials, ShanghaiTech University, Shanghai 201210, China}
\affiliation{Liaoning Academy of Materials, Shenyang 110167, China}

\bibliographystyle{apsrev4-2}

\begin{abstract} 
Charge transfer is a common phenomenon in van der Waals heterostructures with proper work function mismatch, which enables electrostatic gating to control band alignment and interlayer charge distributions. This provides a tunable platform for studying coupled bilayer correlated electronic systems. Here, we theoretically investigate heterostructures of rhombohedral multilayer graphene (RMG) and an insulating substrate with gate-tunable band alignment. We first develop a self-consistent electrostatic theory for layer charge densities incorporating charge transfer, which reproduces the experimentally observed broadened and bent charge neutrality region. When the substrate's band edge has a much larger effective mass than RMG, its carriers may form a Wigner crystal at low densities. This generates a quantum superlattice that induces topological flat bands in the RMG layer through interlayer Coulomb interactions, which may further lead to Chern insulators driven by Coulomb interactions within the RMG layers. Conversely, with comparable effective masses, we find an interlayer excitonic insulator state at charge neutrality stabilized by interlayer Coulomb coupling. Our work establishes these charge-transfer heterostructures as a rich platform for topological and excitonic correlated states.%\sout{, opening an avenue for ``charge-transferonics''}.
\end{abstract}

\maketitle

Two-dimensional (2D) van der Waals heterostructures \cite{VDWhetero-2013}, assembled by stacking chemically distinct atomically thin layers in a designed sequence, provide a highly tunable platform for emergent quantum phenomena including topological insulators, magnetic skyrmion textures and fractional Chern insulators \cite{zhou-trilayer-nature21, skyrmion-tong-nn2018, CI-Huang-np2021, moire-chgr-2020, moire-xiej-arxiv2025, moire-Choi-nature2025, moire-Waters-prx2025, moire-xianghx-2025, moire-Lu-nature2025, moire-Aronson-prx2025, moire-Zhou-nc2024, moire-hanxy-nano2024, moire-Ding-prx2025, moire-zhengJ-2024, moire-wang-2025,hetmd-Regan-2020, hetmd-Tang-2020, hetmd-Xu-2020}. A key ingredient underlying the rich physics is the ability to control the electronic structure in situ with electrostatic gates. If the band edges of the two constituent layers have overlap, electrons would tunnel from an occupied band in one layer into an empty band in the other, which defines a charge-transfer heterostructure. Interlayer Coulomb interactions can then generate synergistic correlated phases that cannot be realized in isolated individual layers \cite{lu-nc23,han-nn2022}. This gate-controlled interlayer charge transfer in 2D van der Waals heterostructures thus opens an avenue for %\sout``charge-transferonics'', i.e., 
engineering novel electronic phases and functionality based on charge-transfer mechanisms \cite{han-nn2022, lu-nc23, han-nc2023, han-nature2024}.

\begin{figure}[htb]
    \centering
    \includegraphics[width=3.5in]{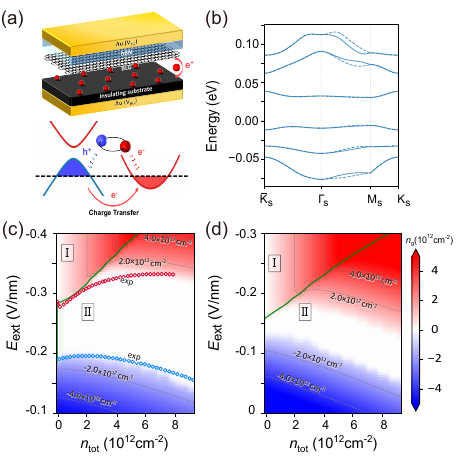}
    \caption{(a) Schematic images of device and the interface electron crystal formed at the surface of substrate (top panel), and an interlayer exciton (bottom panel). (b) Non-interacting energy band of BLG modulated by interlayer superlattice potential. In (c,d), we show the carrier density colormap of BLG ($n_\text{g}$) in our theoretically modelled BLG-CrOCl heterostructure in the parameter space of external vertical electric field ($E_\text{ext}$) and total carrier density ($n_\text{tot}$), with interlayer distance $d_0=4.5$\,\AA\, and $d_0=8$\,\AA, respectively. Regions I and II, separated by solid green line, denote domains without and with charge transfer, respectively. The gray dashed lines mark iso-doping levels of BLG. The red and blue scattered hollow points in (c) denote the experimentally measured charge neutral boundaries on the electron side and hole side. Experimental data is adapted from \cite{han-nc2023} upon the approval of the corresponding author.}
    \label{fig:1}
\end{figure}

A promising realization is provided by stacking graphene multilayers on insulating substrates with proper work function mismatch. Assisted by electrostatic gating, the low-energy bands of graphene can be aligned with the band edges of the substrate, offering a fertile playground for studying the interplay between strong correlations and band topology. As schematically illustrated in the top panel of Fig.~\ref{fig:1}(a), when the charge density induced in the substrate remains below a critical threshold, the transferred electrons can form a Wigner crystal at the interface driven by Coulomb interactions \cite{wigner-pr34, xueqk-WC2026}. This interfacial charge order generates a long-wavelength superlattice potential that acts on graphene via interlayer Coulomb interactions and narrows the subband bandwidth, thereby enhancing electronic correlations, in close analogy to the scenario of moir\'e materials \cite{moire-Andrei2021, BM-model}.   Indeed, charge-transfer-driven correlated states have been experimentally realized in a series of graphene-insulator heterostructures. These states manifest as unusually robust quantum Hall states in monolayer graphene-CrOCl \cite{han-nn2022}, unconventional correlated insulating state in bilayer graphene (BLG)-CrOCl \cite{han-nc2023}, and analogous phenomena in other graphene-based heterostructures \cite{rmg-CrX3-nano22, yankowitz-nano22, gra-tas2-eva-nc2024}.

Compared with monolayer graphene, rhombohedral multilayer graphene (RMG) hosts low-energy bands that are flattened by interlayer hopping and can carry substantial Berry curvature, thereby enhancing interaction effects and facilitating the emergence of topological correlated states \cite{zhou-trilayer-nature21, rmg-sc-Zhou-nature2021,rmg-Shi2020,rmg-Myhro-2018, rmg-Hanth-2023,ju-chern-natnano2023, rmg-hanth-sci2024, rmg-chgr-sci2024, rmg-Hanth-2025,rmg-sc-patterson2024, rmg-sc-yang2025, rmg-chgr-nano2024, rmg-zhy-nano2025, rmg-zhhy-pnas2024,guozhq-nc2025,rmg-berrytrashcan-bernevig2025}. It is therefore natural to expect that even more exotic correlated phases may arise in RMG-insulator charge-transfer heterostructures. In this work, we theoretically investigate charge-transfer heterostructures of RMG on insulating substrates under realistic conditions. We develop a self-consistent electrostatic screening theory that quantitatively reproduces the puzzling experimentally observed bent and broadened charge neutrality region \cite{han-nc2023}. First, separating the electronic degrees of freedom of the two layers \cite{lu-nc23}, we find that a sufficiently heavy substrate band edge's mass enables dilute carriers to form a Wigner crystal, which generates superlattice potential exerting on RMG, thus produces topological flat subbands (see Fig.~\ref{fig:1}(b) for BLG) \cite{cano-bilayer-prl23,cano-multilayer-prb23,zeng-prl24,cano-bGrsuperlat-prb-2024,zhan-patterned-prb25,shi-kagomeFCI-prl-2025}. In this regime, Chern-insulator phases are found at subband fillings $\nu=1,2$ for sufficiently large superlattice period driven by intralayer Coulomb interactions within RMG.

Then, treating both layers on equal footing, we find that opposite charge dopings at overall charge neutrality favors interlayer electron-hole bindings [bottom panel of Fig.~\ref{fig:1}(a)] and stabilize an interlayer excitonic insulator (IEI) under appropriate conditions \cite{EI-zxj-prl1995,EI-pikulin-prl2014,EI-Qzz-sciadv2019,EI-seradjeh-prl2009,EI-shyl-prb2024,EI-zhjh-prb2024,EI-zhjh-prb2025,EI-zyx-prb2020}. With excess charge dopings, the substrate's carriers develop the same Wigner-crystal-like charge order assumed above, supporting the consistency and physical plausibility of the Wigner-superlattice scenario described above.

\paragraph{Electrostatic screening model} 
To accurately determine layer charge density distributions for the charge-transfer heterostructures, one needs to carefully treat the screening of vertical electric field $E_{\text{ext}}$ applied to the system. We start with the non-interacting Hamiltonian $H^{0}= H^{0}_{\text{RMG}}+H^{0}_{\text{sub}}$, where $H^{0}_{\text{RMG}}$ is the non-interacting $\k\cdot\p$ model of RMG \cite{supp_info}.%:
%\begin{equation}
%    H_{\text{sub}}^0 = \sum_{k_s} (\frac{\hbar^2 k_s^2}{2m^*}+E_{\text{CBM}}) \dop^{\dagger}(\bk_s)\dop(\bk_s).
%\end{equation}
%Here $\dop^{(\dagger)}(\bk_s)$ annihilates (creates) an electron with a wave vector $\bk_s$ expanded around the band edge of the substrate. 
Without loss of generality, we suppose the carriers are transferred to the conduction band minimum (CBM) of the substrate, which has a relative energy shift $E_{\text{CBM}}$ with respect to the charge neutrality point of RMG. The interlayer hopping between RMG and the substrate is typically negligible, as confirmed by density functional theory (DFT) calculations \cite{han-nn2022}, because the interlayer hopping is exponentially suppressed due to the large interlayer distance and lattice mismatch in such heterostructures. This endows an emergent layer $U(1)$ symmetry for the low-energy electrons in such system, the spontaneous breaking of which gives rise to an IEI, as will be discussed later.

Under $E_{\text{ext}}$, diagonalizing the above Hamiltonian yields the charge distributions at different layers of the RMG-substrate heterostructure. The redistributed charges, however, modify the interlayer electrostatic potential differences through two distinct and competing mechanisms: one from conventional dielectric screening, and the other from the modification of energy dispersion due to exchange effects, namely the Fock self energy correction. To capture both effects, we formulate a self-consistent screening model incorporating the Fock correction  alongside conventional dielectric screening effects. Specifically, we first calculate the layer-resolved charge densities using the above non-interacting Hamiltonian, which serves as the source term in the Poisson equation of electrostatic potential. Meanwhile, the Fock self-energy correction is incorporated into our screening model as well. \Blue{Since we focus on the RMG charge neutrality region, the dominant Fock correction reduces to a shift of the substrate CBM, $\delta E_{\text{CBM}}=-e^2\sqrt{\pi n_{\text{sub}}}/(4\pi\epsilon_0\epsilon^\text{sub}_\parallel)$, where $n_{\text{sub}}$ is the substrate carrier density and $\epsilon^\text{sub}_\parallel$ is the background relative dielectric constant. This simplified treatment is further justified, in the subsequent discussion of Chern-insulator states in RMG, by the separable-wavefunction, or ``Born-Oppenheimer-type'', approximation \cite{lu-nc23,miao-arxiv26}, whose validity is verified a posteriori by our full-band calculations that explicitly include interlayer exchange effects.} The updated electrostatic potential is then fed back into the Hamiltonian to further calculate layer charge density. This calculation is iterated until both the layer charge density and the electrostatic potential converge. More details are given in Supplemental Materials \cite{supp_info}.

%\Blue{When the substrate's effective mass is much larger than that of RMG, as in BLG-CrOCl, the many-body ground state of the heterostructure is well approximated by a product state of the RMG and the substrate's electrons, the latter of which may form a Wigner crystal at low density. Under this separable-wavefunction, or ``Born-Oppenheimer-type'', approximation \cite{lu-nc23,miao-arxiv26}, which is verified a posteriori by our full-band Hartree-Fock (HF) calculations, the interlayer exchange effect is negligible. In addition, when the substrate carrier density is much larger than the carrier density induced in RMG, as occurs near the RMG charge neutrality region, the exchange self-energy correction within RMG is much smaller than that in the substrate and can be neglected in the electrostatic screening model. Under these two conditions, Under these two conditions, the dominant Fock correction reduces to a shift of the substrate CBM, $\delta E_{\text{CBM}}=-e^2\sqrt{n_{\text{sub}}}/(4\pi\epsilon_0\epsilon_s)$, where $n_{\text{sub}}$ is the substrate carrier density and $\epsilon_s$ is the background relative dielectric constant. [XL: maybe replace it with a shorter version and move the full discussion to SI?]} 

We illustrate our electrostatic screening model using a heterostructure consisted of BLG and CrOCl. \Blue{The same calculations have also been performed for the trilayer graphene-CrOCl heterostructure, with details presented in the Supplemental Material \cite{supp_info}.} The CBM of CrOCl is mostly contributed by Cr's $3d$ orbitals, which thus has a large effective mass $m^*\approx 1.3 m_0$ \cite{lu-nc23,han-nn2022} ($m_0$ is bare electron mass). \Blue{We take $E_{\text{CBM}}=-0.13$\,eV with respect to the Dirac point of graphene, as extracted from DFT calculations \cite{han-nn2022}. Under a vertical electric field, charge carriers are transferred to one surface of the CrOCl substrate, and DFT calculations show that the corresponding Fermi surface is composed of ellipses at low doping levels ($\le 10^{13}$\,cm$^{-2}$) \cite{lu-nc23}. We therefore describe the low-energy electrons at the substrate surface by $H^{0}_{\text{sub}}$, an effective parabolic two-dimensional surface-band model with anisotropic effective masses $m^{*}_x$ and $m^{*}_y$ in the crystalline $x$ and $y$ directions, respectively. We use the isotropic approximation $m^*=\sqrt{m^*_x m^*_y}\approx 1.3 m_0$. The effect of mass anisotropy is analyzed explicitly in the Supplemental Material \cite{supp_info}, where we show that it leads only to minor quantitative changes in the charge-transfer and electrostatic screening results.}

It is worth noting that different choices of $m^{*}$ and antiferromagnetic configurations of CrOCl do not change the qualitative conclusions presented in this work as magnetic exchange coupling between CrOCl and RMG is negligibly weak \cite{han-nn2022,yankowitz-nano22}. Taking BLG for illustration, Fig.~\ref{fig:1}(c,d) present the carrier density of BLG $n_{\text{g}}$ in the parameter space spanned by $E_{\text{ext}}$ and the total carrier density $n_\mathrm{tot}=n_{\text{g}}+n_{\text{sub}}$, for interlayer distance $d_0=4.5$\,\AA\ [Fig.~\ref{fig:1}(c)], and $d_0=8$\,\AA\ [Fig.~\ref{fig:1}(d)], respectively. These phase diagrams exhibit a common feature: upon tuning $E_\mathrm{ext}$ and $n_\mathrm{tot}$, the graphene system evolves from a conventional region (Region \I), in which the substrate remains undoped, to a region with significant charge transfer (Region \II). Within Region \II, increasing the amplitude of $E_\mathrm{ext}$ drives BLG continuously from hole-doped state (blue), through the broad and bent charge neutrality region (white), to electron-doped state (red). The iso-density levels in BLG, shown as grey dashed curves, exhibit sharp turning points, highlighting the nonlinear dependence of the equilibrium carrier density on both $E_\mathrm{ext}$ and $n_\mathrm{tot}$ due to charge transfer.

\Blue{Notably, our calculations for $d_0=4.5$\,\AA\ quantitatively reproduce the pronounced broad and bent charge region of BLG that has been observed in previous transport measurements of BLG-CrOCl devices \cite{han-nc2023}. The experimental boundaries of charge neutral region have been presented as red and blue hollow points in \ref{fig:1}(c).} These characteristic features of RMG charge-transfer heterostructure are attributed to the competition between two different mechanisms. On the one hand, electrostatic screening associated with the redistribution of charge between the substrate and RMG reduces the electric field between the two subsystems by $n_{\mathrm{sub}}e^2/(\epsilon_0\epsilon_s)$. This screening effect suppresses further charge transfer into the substrate and, by itself, would lead to charge-neutrality boundary lines that decrease approximately linearly with $n_{\mathrm{sub}}$. On the other hand, a negative Fock self-energy for the transferred carriers in the substrate lowers the substrate CBM and thereby favors additional charge transfer into the substrate. At low $n_{\mathrm{sub}}$, this Fock correction, which scales as $\sqrt{n_{\mathrm{sub}}}$, dominates over the linear electrostatic screening contribution. This competition gives rise to upward-curved iso-density lines and bent phase boundaries separating the hole-doped or electron-doped regions from the charge-neutral region. As $n_{\mathrm{sub}}$ increases, the linear electrostatic screening term becomes dominant, causing the iso-density lines to decrease almost linearly. The same competition also accounts for the pronounced broadening of the charge-neutrality region \cite{supp_info}. \Blue{The Fock correction should be viewed as a minimal low-energy approximation. In realistic CrOCl interfaces, disorder, finite thickness, and localization effects may renormalize its prefactor or modify its precise density dependence. However, our screening mechanism does not rely on the exact square-root form. The key ingredient is the competition between a sublinear interaction-induced correction from heavy substrate carriers and the linear electrostatic charging contribution. These non-ideal effects may change quantitative parameters, such as the effective Fock-shift coefficient and density scale, but do not alter the qualitative bending and broadening mechanism of the charge-neutrality region.} 

\Blue{Furthermore, as shown in Fig.~\ref{fig:1}(d), our theoretical results with increased interlayer distance $d_0=8\,$\AA\ indicate a drastically reduced charge neutrality region of BLG, which is in agreement with experimental observation \cite{han-nc2023}. The agreement between our theoretical predictions and experimental observations justifies our modelling for such charge-transfer heterostructure \cite{han-nc2023,supp_info}.} 

% \sout{This reduction originates from the fact that $d_0$ affects both the interlayer potential difference and the dielectric screening effect. }
%\sout{At the lower boundary of the CNP, the Fermi level aligns with the valence band maximum (VBM) of RMG, while at the upper boundary it aligns with CBM of RMG. When other parameters are fixed, increasing $d_0$ reduces the magnitude of external electric field corresponding to the lower boundary and also diminishes the difference in the magnitude of external electric fields between the upper and lower boundaries. Moreover, as the screening effect becomes stronger with increasing $d_0$, the inflection point where the phase boundary changes from upward-curving (dominated by the Fock correction) to downward-curving (dominated by the dielectric screening effect) shifts leftward. For a detailed comparison of the phase boundaries as a function of $d_0$, see the Supplemental Material \cite{supp_info}.}
 %\Blue{[XL: we can add a subsection in SI to discuss the quantitative agreement with Han's work on BLG-CrOCl.]}. %More importantly, our approach captures the essential physics governing interlayer charge redistribution and dielectric screening, thereby establishing a reliable foundation for understanding electron-electron interaction effects in graphene-based charge-transfer heterostructures. 

\paragraph{Chern insulator boosted by Wigner crystallization}

The full interacting Hamiltonian of the heterostructure can be decomposed into three parts: the RMG part, the substrate part, and the interlayer Coulomb coupling between them \cite{lu-nc23}. The large effective mass of the substrate's carriers allows for the ``Born-Oppenheimer-type'' treatment to the heterostructure, i.e., the ground-state wavefunction is expressed as the product state of that of RMG and that of the substrate, as discussed above \cite{lu-nc23}. As long as $n_\mathrm{sub}$ is below a threshold value,  the corresponding Wigner-Seitz radius $r_s \gtrapprox 31$ \cite{needs-wigner-qmc-prl09}, a Wigner-crystal-like long-wavelength charge ordered state is expected to form at the surface of the substrate \cite{xueqk-WC2026}. This imposes an effective superlattice potential to the electrons in the RMG layer. Thus, we obtain the low-energy effective continuum model for the RMG coupled to insulating substrate \cite{supp_info}
\begin{equation}
    \hat{H}_{\text{RMG}}^{0, \mu} = \hat{H}_{\text{RMG}}^{0, \mu} + U_s(\br) + U_d
    \label{eq:Hrmg}
\end{equation}
where $U_d$ is the layer-resolved on-site energy obtained from self-consistent electrostatic screening calculation. $U_s(\br)$ is the superlattice potential from substrate's long-wavelength charge order with the period $U_s(\br) = U_s(\br+\bf{L_s})$ with $\bf{L_s}$ superlattice vector. %Such a continuum model is adopted throughout this section given that $L_s \ge a$ ( $a=2.46$\,\AA \ is the lattice constant of graphene) is always fulfilled with low carrier density of the substrate. More details about the full many-body Hamiltonian are given in Supplemental Materials \Blue{[XL: cite Supp. Mat.]}.

\begin{figure}[htb!]
    \centering
    \includegraphics[width=3.5in]{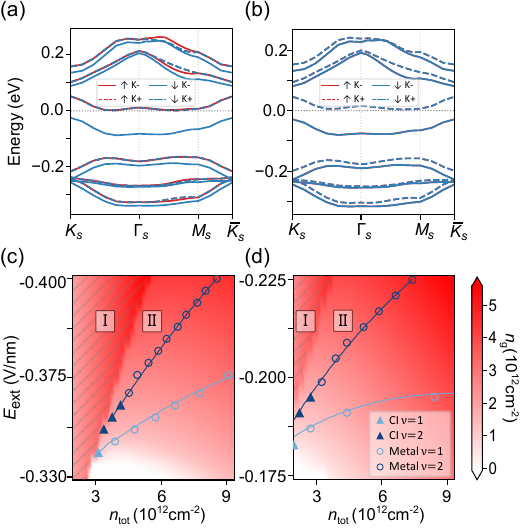}
    \caption{HF single particle spectra for HF ground state of BLG: (a) at $E_\text{ext}=-0.3417 $\,$\text{V}/\text{nm}$ and $n_\text{tot}= 3.14\times10^{12}$\,cm$^{-2}$ with filling factor $\nu=1$; and (b) at $E_\text{ext}=-0.3542 $\,$\text{V}/\text{nm}$ and $n_\text{tot}= 3.78\times10^{12}$\,cm$^{-2}$ with filling factor $\nu=2$. The HF ground states along iso-filling lines $\nu=1$ and $\nu=2$ are shown in (c) for BLG-CrOCl with $d_0=4.5$\,\AA, and in (d) for BLG-CrOCl with $d_0=8$\,\AA, where color coding represents carrier density $n_\text{g}$ of RMG, ``CI'' denotes Chern insulator state.} 
    \label{fig:3}
\end{figure}

Considering the continuum model with renormalized parameters due to the Coulomb potential from high-energy electrons \cite{gonzalez_nuclphysb1993, kang-rg-prl20, guo-prb24, supp_info}, we find that for a Wigner superlattice period exceeding 10\,nm, isolated flat bands appear. The lowest conduction band (LCB) carries a nonzero Chern number ($|C|=1$) as shown Fig.~\ref{fig:1}(b) for BLG \cite{supp_info}. In Fig.~\ref{fig:3}(c) and (d), we present the results for Wigner-crystal-coupled BLG with $d_0=4.5\,$\AA\ and $d_0=8\,$\AA,  respectively. The gray shaded area denotes the conventional state without charge transfer, while dashed lines mark iso-density levels in the BLG layer. Solid lines indicate the iso-filling curves $\nu=1$ and $\nu=2$ of the Wigner-crystal supercell ($\nu = \sqrt{3}L_s^2 n_{\text{g}}/2$, assuming a triangular superlattice as expected for a Wigner crystal \cite{wigner-pr34}). These topologically nontrivial flat bands, when filled with electrons, are expected to host novel quantum states \cite{cano-bilayer-prl23,cano-multilayer-prb23,zeng-prl24,cano-bGrsuperlat-prb-2024,zhan-patterned-prb25,shi-kagomeFCI-prl-2025}.

To explore the many-body ground state of the RMG, we further study the $e$-$e$ interaction effects using a band-projected Hartree-Fock (HF) approach \cite{supp_info}. Taking BLG for illustration, at integer fillings $\nu=1, 2$, Chern insulator (CI) states emerge as HF ground states through spontaneous time-reversal symmetry breaking. The typical HF band structures of the Chern insulators are  shown in Fig.~\ref{fig:3}(a) and (b), for $\nu=1$ and $\nu=2$, with $d_0=4.5\,$\AA. While a spin-valley polarized $|C|=1$ CI emerges as the ground state for $\nu=1$, a spin-degenerate but valley-polarized $|C|=2$ Chern insulator is found at $\nu=2$. \Blue{The lifting of spin and/or valley degeneracy is solely driven electron-electron Coulomb interactions through spontaneous symmetry breaking.} Similar correlated Chern insulators are also identified for enlarged interlayer distance $d_0=8$\,\AA\ \cite{supp_info}. \Blue{To examine possible competing density-wave orders, we further compute the generalized static susceptibility within the random-phase approximation. No divergent susceptibility is found at a representative point inside the $\nu=1$ CI phase, suggesting that the CI state is not preempted by a trivial density-wave instability in this regime \cite{supp_info}. While the CI states persist up upon varying $d_0$, they still only emerge in a small region, and are not robust to the increase of dielectric constant \cite{supp_info}. Similarly, the Chern insulator phase persists in the trilayer graphene-CrOCl heterostructure, as detailed in the Supplemental Material \cite{supp_info}.}

In Fig.~\ref{fig:3}(c) and (d), we present the calculated ground states of RMG along the $\nu=1$ and $\nu=2$ iso-filling lines, where open circles denote metallic states and filled triangles indicate CI states. CI states arise near the charge-transfer phase boundary, where the formed Wigner superlattice exhibits a long period due to the low transferred carrier density. In this regime, the subbands exhibit narrow bandwidths, leading to the formation of flat bands with nonzero Chern numbers after coupled with Wigner superlattice potential, which subsequently evolve into Chern insulators driven by interactions. Away from the charge-transfer boundary, the superlattice period shortens, resulting in more dispersive subbands that favor a metallic state rather than a gapped insulator.

\paragraph{Interlayer excitonic insulator}

As discussed above, the full interacting Hamiltonian of the coupled bilayer system possesses an emergent layer $U(1)$ symmetry due to the negligible interlayer hopping. Nevertheless, charge carriers of the two layers are coupled via interlayer Coulomb interactions. If the two layers are oppositely charged, the electron-hole pairs from the two layers may bind together driven by interlayer Coulomb interactions and form a condensate, leading to an excitonic insulator state. Such a state is  beyond the ``Born-Oppenheimer-type'' treatment discussed above. We need to consider the RMG-substrate heterostructure as a unified quantum many-body system. To this end, we adopt a full-band HF method \cite{supp_info} to study the ground state of the heterostructure. Specifically, we treat the effective mass of substrate's carriers $m^*$ and initial $E_{\text{CBM}}$ as two parameters, and vary them to explore the ground-state phase diagram.

\begin{figure*}[htb!]
    \centering
    \includegraphics[width=7in]{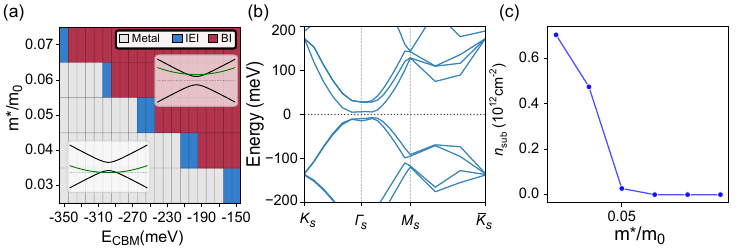}
    \caption{(a) HF ground state phase diagram of BLG-substrate heterostructure. Insets at lower left corner and upper right corner schematically show the band structures of metallic state and band insulator state, respectively. Green lines and black lines denote substrate's and BLG's bands, respectively. (b) HF single particle spectrum of IEI state with $E_{\text{CBM}} = -260$\,meV and effective mass $m^*=0.05m_0$. (c) Substrate's carrier density $n_\mathrm{sub}$ as a function of $m^*/m_0$ with $E_\mathrm{CBM}^0=-250$\,meV.}
    \label{fig:4}
\end{figure*}

Taking BLG for illustration, Fig.~\ref{fig:4}(a) presents the phase diagram of the BLG-substrate heterostructure. At large $|E_\mathrm{CBM}|$ and small $m^*/m_0$, the system favors metallic state, corresponding to the gray region. The inset at the left lower corner in Fig.~\ref{fig:4}(a) schematically illustrates the band alignment that leads to this metallic state, where the horizontal dotted line marks the Fermi level. While for smaller $|E_\mathrm{CBM}|$ and large $m^*/m_0$, the substrate's CBM is pushed above the Fermi level such that charge transfer from BLG to the substrate is suppressed. Hence, the system behaves as a conventional isolated BLG. This regime is identified as a band insulator (``BI'' in Fig.~\ref{fig:4}(a)) state and marked in red. We schematically illustrate the band alignment that lead to the band insulator state in the inset at the right upper corner of Fig.~\ref{fig:4}(a). Between these two states, a narrow region of IEI state (blue region), characterized by spontaneous layer $U(1)$ symmetry breaking, emerges along the diagonal from the upper left to the lower right. In this phase, the system exhibits a finite gap at the Fermi level due to the interlayer binding of electron-hole pair. The calculated HF band structures of IEI state is presented in Fig.~\ref{fig:4}(b), with parameters $E_{\text{CBM}}=-260$\,meV and $m^*/m_0 =0.05$, featuring a pronounced gap at chemical potential.

The restricted parameter-space window for the emergence of the IEI state can be understood as follows. The electrons on the substrate and the holes from RMG are bound together via Coulomb attraction, forming interlayer exciton with hydrogen-like orbital characteristics. The reduced mass $\mu$ of this interlayer exciton is $1/\mu = 1/m^*+1/2m^*_\text{g}$, where $m^*_{\text{g}}$ is the effective mass of the valence band maximum of RMG and the factor of 2 comes from the valley degeneracy. The corresponding binding energy $E \propto -\mu/\epsilon_r^2$, with effective Bohr radius $\sim \epsilon_r (m_0/\mu) a_{\text{B}}$, where $a_{\text{B}}$ is Bohr radius of hydrogen atom. Consequently, the interlayer exciton bound state is favored when the reduce mass $\mu$ is large. On the other hand, the condensation energy of IEI (in the low exciton density limit) depends on both the reduced mass and $n_\mathrm{sub}$. However, this coupled heterostructure exhibits a distinct charge transfer mechanism: as shown in the Fig.~\ref{fig:4}(c), fixing $E_{\text{CBM}}=-250\,\,$meV, the carrier density transferred to substrate $n_\mathrm{sub}$, determining the maximal density of exciton, decreases with the increase of $m^*$ due to the screening effect and Fock energy corrections. $n_\mathrm{sub}$ eventually drops to 0 for large $m^*$, with the disapperance of charge transfer. %In contrast, under charge neutrality condition, when $m^*$ decreases, $n_\mathrm{sub}$ increases from 0 to a finite value, leaving equally amount of holes on RMG. 
Therefore, with the decrease of $m^*$,  a phase transition from the band insulator to IEI is triggered as $n_{\text{sub}}$ increases from 0 to finite value. If the effective mass $m^*$ is even smaller with larger $n_{\text{sub}}$, kinetic energy would dominate over Coulomb interaction energy, and the system becomes metallic. %\Blue{This physical picture is consistent with an analysis by fixing $m^*$ which shows that} the carrier density transferred to substrate $n_\mathrm{sub}$ increases when $\vert E_\mathrm{CBM}^0\vert$ increases, leading to the similar phase transition from band insulator to IEI, then to metallic state. 
We note that, if the whole system is electron or hole-doped, Wigner crystal state can be obtained at the substrate side for large $m^*$ and/or low $n_{\text{sub}}$ using our full-band HF approach, without imposing the Born-Oppenheimer-type treatment or the proposed screening model. This provides strong self-consistent support and corroborates the physical picture underlying the earlier Born-oppenheimer-type treatment.

In conclusion, our work establishes a comprehensive theoretical framework for RMG-insulator charge-transfer heterostructures. Starting with a self-consistent electrostatic screening model, we unveil the competition mechanism that leads to the puzzling experimental transport features, i.e., the bent and broadened charge neutrality region. We further study the correlated and topological properties of the charge-transfer heterostructures in different regimes. When the effective mass of substrate's carriers is much larger than that of RMG, the transferred carriers at the substrate may form a Wigner crystal, which imposes a superlattice electrostatic potential to the low-energy electrons in RMG through interlayer Coulomb interactions. This suppresses the kinetic energy, and endows nontrivial topological properties to the low-energy carriers in RMG. As a result, Chern insulators emerge at integer fillings of the Wigner supercell driven by long-range electron-electron interactions. When the effective mass of substrate's carriers is comparable to that of RMG, interlayer excitonic insulator emerges at charge neutrality. Our work thus provides a unified theoretical framework that bridges charge reconstruction, topological properties and strong correlations in charge-transfer van der Waals heterostructures.

\acknowledgements 
This work is supported by National Natural Science Foundation of China (grant no. 12404221, no. U23A6002 and no. 12550403), the National Key Research and Development Program of China (grant no. 2024YFA1410400 and no. 2022YFA1604400/03), the Strategic Priority Research Program of Chinese Academy of Sciences (grant no. XDB1710000), and Shanghai Science and Technology Innovation Action Plan (grant no. 24LZ1401100).

\paragraph{Data Availability.} The data needed to reproduce the figures in this article are available in \cite{Shiyr2026}.

\widetext
\clearpage

\makeatletter
\def\@fnsymbol#1{\ensuremath{\ifcase#1\or \dagger\or \ddagger\or
		\mathsection\or \mathparagraph\or \|\or **\or \dagger\dagger
		\or \ddagger\ddagger \else\@ctrerr\fi}}
\makeatother

\begin{center}
\textbf{\large Supplemental Materials for ``Correlated states in charge-transfer heterostructures based on rhombohedral multilayer graphene"} \\
\vspace{0.5cm}
Yanran Shi, Min Li, Xin Lu \footnote{lvxin@shanghaitech.edu.cn} and Jianpeng Liu \footnote{liujp@shanghaitech.edu.cn}
\end{center}

\setcounter{equation}{0}
\setcounter{figure}{0}
\setcounter{table}{0}
\setcounter{section}{0}
\makeatletter
\renewcommand{\theequation}{S\arabic{equation}}
\renewcommand{\thesection}{S\arabic{section}}
%\renewcommand{\thefigure}{S\arabic{figure}}
%\renewcommand{\bibnumfmt}[1]{[S#1]}
%\renewcommand{\citenumfont}[1]{S#1}
%%%%%%%%%% Prefix a "S" to all equations, figures, tables and reset the counter %%%%%%%%%%
%\newcommand{\setlabel}[1]{\edef\@currentlabel{#1}\label}
%\captionsetup{justification=Justified, singlelinecheck=false}
\renewcommand{\figurename}{Supplementary Figure}
\renewcommand{\tablename}{Supplementary Table}

\def\bibsection{\section*{References}} % remove the ugly line before bibliography   
\tableofcontents

\section{$\bk \cdot \p$ model of rhombohedral multilayer graphene}
\label{sec:S1}
The charge-transfer heterostructures based on rhombohedral multilayer graphene (RMG) described in the main text can always be divided into three parts: the RMG part, the substrate part and the interlayer Coulomb coupling between them. In this section, we present a detailed derivation of the Hamiltonian for the $\bk \cdot \p$ model of RMG. Since we focus on the low energy physics, the non-interacting Hamiltonian of RMG can be described by a general effective continuum model directly derived from the atomistic tight-binding Hamiltonian. 

We define the primitive lattice vectors of graphene unit cell as $\a_1=a(1,0)$ and $\a_2=a(1/2, \sqrt{3}/2)$ with $a=2.46\,\text{\AA}$. The corresponding reciprocal lattices are constructed as $\b_1=2\pi/a (1,-1/\sqrt{3})$ and $\b_2=2\pi/a (0,2/\sqrt{3})$ from $\a_i \cdot \b_j=2\pi\delta_{ij}$. The two sublattices forming honeycomb lattice are seated at $\bm{\tau}_a=a(0,-1/\sqrt{3})$ and $\bm{\tau}_b=(0,0)$, respectively. We consider the situation that the next layer is shifted in the in-plane direction $a(0,-1/\sqrt{3})$ with respect to the previous layer, which defines the stacking chirality. The Dirac points of monolayer graphene are located at $\mathbf{K}_+= -4\pi/3a(1,0)$ and $\mathbf{K}_-=4\pi/3a(1,0)$. Then we derive the $\k \cdot \p$ model from the atomistic Slater-Koster tight-binding model based on carbon's $p_z$-like Wannier orbitals:
\begin{equation}
    H^0_{\rm{RMG}}=\sum_{i l\alpha,j l^{\prime}\alpha^{\prime}}
    {-t\left ( 
    \R_i+\bm{\tau}_{\alpha}+l d_0 \mathbf{e}_z
    -\R_j-\bm{\tau}_{\alpha^{\prime}}-l^{\prime} d_0 \mathbf{e}_z 
    \right )} 
    \hat{c}^{\dagger}_{i l \alpha}\hat{c}_{j l^{\prime}\alpha^{\prime}}\;,
\end{equation}
where $i$, $j$ represents lattice sites and $\R_i$, $\R_j$ represents lattice vectors in graphene. $l$ and $l^{\prime}$ are layer indexes while $\alpha$ and $\alpha^{\prime}$ are sublattice indexes. $d_0$ is the interlayer distance and $\mathbf{e}_z$ is a unit vector along out-of-plane direction. $t(\mathbf{d})$ is hopping amplitude between two $p_z$ orbitals displaced by vector $\mathbf{d}$, which is expressed in the form:
\begin{align}
    -t(\bd) &= 
    V_{pp\pi}\left [ 1-\left ( \frac{\bd \cdot \mathbf{e}_z}{d} \right )^2  \right ] +V_{pp\sigma}\left ( \frac{\bd \cdot \mathbf{e}_z}{d} \right )^2 \\
    \begin{split}
        V_{pp\pi}&=V_{pp\pi}^0\exp\left ( -\frac{\left | \bd \right |-a/\sqrt{3}}{r_0}  \right ) \\
        V_{pp\sigma}&=V_{pp\sigma}^0\exp\left ( -\frac{\left | \bd \right |-d_0}{r_0}  \right ) 
    \end{split}
\end{align}
where $V_{pp\pi}^0=-2.7\, \eV$, $V_{pp\sigma}^0=0.48\, \eV$ and $r_0=0.184a$ \cite{moon-tbg-prb13,Moon-Koshino-prb2014}.

In our model, we consider only interlayer hopping between two nearest graphene sheets. After Fourier transformation to $\k$ space, and expand the (Fourier transformed) tight-binding model in the vicinity of Dirac points, we obtain the $\k\cdot\p$ model of RMG, with the intralayer and interlayer parts of Hamiltonian expressed as
\begin{align}
    h_{\rm{intra}}^{0,\mu}
    &= -\hbar v_F^0 \k \cdot \bm{\sigma} ^{\mu} \\
    h_{\rm{inter}}^{0,\mu}
    &=
    \begin{pmatrix}
        \hbar v_{\perp}(\mu k_x+ik_y) & t_{\perp} \\
        \hbar v_{\perp}(\mu k_x-ik_y) & \hbar v_{\perp}(\mu k_x+ik_y)
    \end{pmatrix}   
\end{align}
$\mu=\pm$ refers to valley index. $\k$ is the wave vector expanded around $\mathbf{K}_+/\mathbf{K}_-$ point. $\bm{\sigma}^\mu=(\mu \sigma _x,\sigma _y )$ are Pauli matrices defined in sublattice space. In term of Slater-Koster transfer integral form, $\hbar v_F^0=5.253\,\eVA$ is the non-interacting band velocity of Dirac fermions in monolayer graphene. $\hbar v_\perp=0.335\,\eVA$  and $t_\perp=0.34\,\eVA$ are Slater-Koster hopping parameters. The low-energy Hamiltonian of $n$-layer RMG of valley $\mu$, denoted as $H_\text{RMG}^{0,\mu}$, is just consisted of the intralayer term $h_{\rm{intra}}^{0,\mu}$ appearing in the diagonal block (in layer space) and the interlayer term $h_{\rm{inter}}^{0,\mu}$ coupling adjacent layers.

%%%%%
\section{\Blue{Fock energy correction}}
\label{sec:S2}
We continue to consider the Fock self-energy correction to electrons transferred from the RMG to the substrate. In the setup described in the main text, we consider the situation that the conduction band minimum (CBM) of the substrate is charge doped. The DFT calculations \cite{lu-nc23} have shown that charges accumulate on the substrate's lower-energy side under a vertical electric field and the Fermi surface is composed of ellipses at extremely low doping ($\le 10^{13}$\,cm$^{-2}$). Hence, $H^{0}_{\text{sub}}$ describes the low-energy interface electrons as a parabolic two-dimensional electron gas with density-of-states effective mass $m^*=\sqrt{m^*_x m^*_y}$, where $m^*_x$ and $m^*_y$ are the principal effective masses at CBM along two orthogonal in-plane directions of the Brillouin zone. The corresponding Hamiltonian reads
\begin{equation}
    H_\text{sub}^0 = \sum_{k_s} (\frac{\hbar^2 k_{s,x}^2}{2m^*_x}+\frac{\hbar^2 k_{s,y}^2}{2m^*_y}+E_{\text{CBM}}) \dop^{\dagger}(\bk_s)\dop(\bk_s).
\end{equation}
In the isotropic limit where $m^*=m^*_x=m^*_y$, the Hamiltonian simplifies to
\begin{equation}
    H_\text{sub}^0 = \sum_{k_s} (\frac{\hbar^2 k_{s}^2}{2m^*}+E_{\text{CBM}}) \dop^{\dagger}(\bk_s)\dop(\bk_s),
\end{equation}
where $k_s^2 = k_{s,x}^2+k_{s,y}^2$. The operator $\dop(\bk_s)$ ($\dop^\dagger(\bk_s)$) annihilates (creates) an electron with a wave vector $\bk_s$ expanded around the CBM within the atomic unit cell of substrate. The $E_{\text{CBM}}$ is the relative energy shift of CBM with respect to the charge neutrality point of RMG. For example, according to the DFT calculations, the value of $E_{\text{CBM}}$ is $-0.13$\,eV for CrOCl. However, long-ranged Coulomb interactions among electrons transferred to the substrate will modify the energy dispersion and consequently change the value of $E_{\text{CBM}}$ \cite{2DEG-prl2009, MC-2DEG-1994}. 

The long-ranged Coulomb interactions between electrons can be written as:
\begin{equation}
    \hat{H}_{\text{int}} = \frac{1}{2S}\sum_{\bk_s, \bk_s', \q_s}\sum_{\sigma,\sigma'} V(\q_s)\hat{d}^\dagger_{\sigma,\bk_s+\q_s}\hat{d}^\dagger_{\sigma',\bk_s'-\q_s}\hat{d}_{\sigma',\bk_s'}\hat{d}_{\sigma,\bk_s}.
\end{equation}
In the Hartree-Fock (HF) approximation, the electron-electron interaction is decoupled at the mean-field level into the Hartree term and the Fock term: 
\begin{equation}
    \hat{H}_{\text{int}} = \frac{1}{S}\sum_{\bk_s\sigma,\bk_s'\sigma'} \left [ V(0) - \delta_{\sigma\sigma'}V(\bk_s'-\bk_s) \right ] n_{\bk_s'\sigma'}\hat{d}^\dagger_{\bk_s\sigma}\hat{d}_{\bk_s\sigma}.
\end{equation}
The Hartree term represents the classical electrostatic potential generated by the average charge density and the Fock term encodes the quantum-mechanical exchange effects. We subsume the Hartree term into the layer-dependent electrostatic energy from the self-consistent screening model introduced in the next section and thus neglected here. Only the Fock term is retained, as it constitutes the primary correction of $E_{\text{CBM}}$. The Fock energy reads:
\begin{equation}
    \begin{aligned}
        V^{\text{x}}_{\bk_s\sigma} &=-\frac{1}{S}\sum_{\bk_s'\sigma'}\delta_{\sigma\sigma'}V(\bk_s'-\bk_s) n_{\bk_s'\sigma'} \\
        & = -\int{\frac{d^2\bk_s'}{4\pi^2}} \frac{e^2}{2\epsilon_0\epsilon_{\parallel}^{\text{sub}}\cdot\frac{1}{|\bk_s'-\bk_s|}n_{\bk_s'\sigma}} \\
    \end{aligned}
\end{equation}
$V(\bk_s'-\bk_s) = e^2/2\epsilon_0\epsilon^\text{sub}_{\parallel}|\bk_s'-\bk_s|$ is the Fourier transformation of 2D long-range Coulomb potential $e^2/4\pi\epsilon_0\epsilon^\text{sub}_\parallel$. $\epsilon_\parallel^\text{sub}$ describes the screening of intralayer Coulomb interactions within the substrate surface state. $S$ is the total area of the system. In the second line of the above derivation, we smear the sum over $\bk_s'$ by replacing it with an integral. Since we focus on the low energy electrons around the CBM of the substrate, we set $\bk_s=0$. We then define $a_F^2 = 2m^*_0\epsilon_F/\hbar^2$ and $b_F^2 = 2m^*_1\epsilon_F/\hbar^2$, where $\epsilon_F$ is the Fermi energy. The above equation can be rewritten as :
\begin{equation}
    \begin{aligned}
        V^{\text{x}}_{\bk_s=0, \sigma}= - \int_{\frac{{k_{s,x}'}^2}{a_F^2}+ \frac{{k_{s, y}'}^2}{b_F^2} \le 1} \frac{d^2\bk_s'}{4\pi^2}\frac{e^2}{2\epsilon_0\epsilon_\parallel^\text{sub}}\cdot\frac{1}{\sqrt{{k_{s,x}'}^2+{k_{s,y}'}^2}}
    \end{aligned}
    \label{eq:Fock_energy}
\end{equation}
To evaluate this integral, we introduce the parameters $t \in [0, 1]$ and $\theta \in [0, 2\pi]$ via the transformation $k_{s,x}'=ta_F\cos\theta$ and $k_{s,y}'=tb_F\sin\theta$ (assuming $a_F > b_F$). The Jacobian determinant of this transformation is given by $J=\frac{\partial(x, y)}{\partial(t, \theta)}=a_Fb_Ft$, substituting into \ref{eq:Fock_energy} yields
\begin{equation} 
    \begin{aligned}
        V^{\text{x}}_{\bk_s=0, \sigma} &= - \int_0^1 dt \int_0^{2\pi} d\theta \frac{a_Fb_Ft}{4\pi^2}\frac{e^2}{2\epsilon_0\epsilon_\parallel^\text{sub}}\cdot\frac{1}{\sqrt{t^2 a_F^2{\cos\theta}^2+t^2 b_F^2 {\sin\theta}^2}} \\
        &= - \int_0^1 dt \int_0^{2\pi} d\theta \frac{a_Fb_Ft}{4\pi^2}\frac{e^2}{2\epsilon_0\epsilon_\parallel^\text{sub}}\cdot\frac{1}{ta_F\sqrt{1 - {\sin\theta}^2+\frac{b_F^2}{a_F^2} {\sin\theta}^2}} \\
        &= - 4b_F\int_0^1 dt \int_0^{\pi/2} d\theta \frac{1}{4\pi^2}\frac{e^2}{2\epsilon_0\epsilon_\parallel\text{sub}}\cdot\frac{1}{\sqrt{1 - \eta^2{\sin\theta}^2}} \\
        &= - 4b_F \frac{1}{4\pi^2}\frac{e^2}{2\epsilon_0\epsilon_\parallel^\text{sub}}K(\eta) \\
    \end{aligned}  
\end{equation}
where $\eta = \sqrt{1 - b_F^2/a_F^2}$ represents the eccentricity. $K(\eta) = \int_0^{\pi/2} d\theta \cdot\frac{1}{\sqrt{1 - \eta^2{\sin\theta}^2}}$ is exactly the complete elliptic integral of the first kind. 

Now we introduce the mass anisotropy ratio $\gamma_m = \frac{min(m^*_0, m^*_1)}{max(m^*_0, m^*_1)}$, so that $\eta=\sqrt{1-\gamma_m}$. Throughout this work, we consider the situation that electrons transferred to the substrate are spin degenerate. Taking the spin degeneracy into account, the Fermi wave vector component along the $y$-direction is given by $b_F =\sqrt{\pi n\sqrt{\gamma_m}}$, where $n$ is the total electron density. Substituting this into the Fock energy expression yields:
\begin{equation}
    V^{\text{x}}_{\bk_s=\mathbf{0}} = - \sqrt{\pi n\sqrt{\gamma_m}}\frac{e^2}{2\pi^2\epsilon_0\epsilon_\parallel^\text{sub}}K(\sqrt{1-\gamma_m}).
    \label{eq:fock_dE_aniso}
\end{equation}
In the isotropic limit $\gamma_m=1$, the complete elliptic integral of the first kind $K(\sqrt{1-\gamma_m})$ becomes $K(0)=\pi/2$ and the Fock energy becomes
\begin{equation}
    V^{\text{x}}_{\bk_s=\mathbf{0}} = - \frac{e^2}{4\pi\epsilon_0\epsilon_\parallel^\text{sub}}\sqrt{\pi n}.
    \label{eq:fock_dE_iso}
\end{equation}
It is exactly the expression of Fock energy correction $\delta E_{\text{CBM}} = -\frac{e^2}{4\pi \epsilon_0 \epsilon_\parallel^\text{sub}} \sqrt{\pi n_{\text{sub}}}$ mentioned in the main text.

\begin{figure}[bth!]
%\begin{center}
    \includegraphics[width=5in]{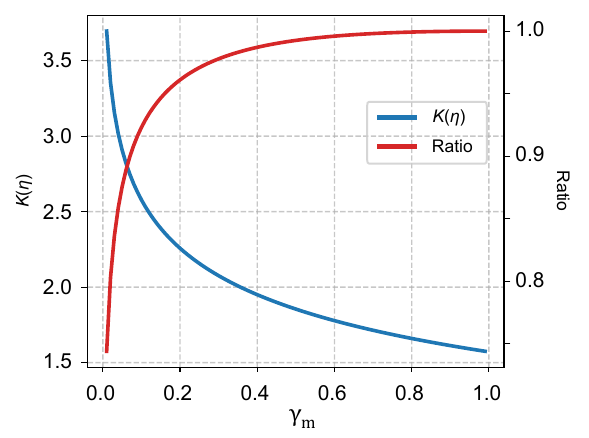}
\caption{Dependence of the complete elliptic integral of the first kind $K(\sqrt{1-\gamma_m})$ (blue curve) and the ratio of the anisotropic Fock energy to its isotropic limit $ V^{\text{x}}_{\bk_s=\mathbf{0}}(\gamma_m)/ V^{\text{x}}_{\bk_s=\mathbf{0}}(1)$ (red curve) on the mass anisotropy ratio $\gamma_m$.} 
%\end{center}
\label{SI_1}
\end{figure}

To quantify the effect of mass anisotropy, we present the dependence of the relevant quantities on the mass anisotropy ratio $\gamma_m$ in \figurename~\ref{SI_1}.  The blue curve shows the behavior of the complete elliptic integral of the first kind $K(\sqrt{1-\gamma_m})$, which increases monotonically as $\gamma_m$ decreases from 1 to 0. The red curve displays the ratio of the Fock energy in the anisotropic case to that in the isotropic limit,i.e., $ V^{\text{x}}_{\bk_s=\mathbf{0}}(\gamma_m)/ V^{\text{x}}_{\bk_s=\mathbf{0}}(1)$. For the case $\gamma_m=0.1$, this ratio is approximately 0.9, indicating a reduction of about 10\% compared to the isotropic case. As \figurename~\ref{SI_3} shown, we have calculated the charge-transfer phase diagrams for both the anisotropic and isotropic models. The resulting charge neutrality regions are found to be almost identical. Based on this observation, we conclude that the isotropic approximation is justified without sacrificing the accuracy of the calculated phase boundaries. Therefore, in practical calculations we adopt the isotropic approximation, which greatly simplifies the analysis while preserving the essential physics.

\section{Self-consistent screening model}
\label{sec: S3}
It has been well known that an externally applied out-of-plane electric field experiences significant screening thanks to the redistribution of electrons across different layers in multilayer graphene system \cite{min-prb23, mccann-bGrEfield-prb-2006, avetisyan-multiGrEfield1-prb-2009, avetisyan-multiGrEfield2-prb-2009, koshino-triGrEfield-prb-2009, guozhq-fci-prb2024}. In this section, we develop a self-consistent screening model for charge transfer heterostructures based on RMG that integrates the self-consistent Hartree screening method described in Ref.~\cite{guozhq-fci-prb2024} to account for the conventional screening effect along with the Fock energy correction to $E_\text{CBM}$ discussed in the above section. This framework enables the determination of the layer-resolved charge density distribution and relative electrostatic energy of graphene layers with respect to the surface of the substrate.

Under an out-of-plane electric field $E_{\text{ext}}$, each graphene layer gets a relative electrostatic energy shift with respect to the substrate, which is expressed as $U_d = e E_{\text{ext}} (d_0+ld_{\text{g}})$ and added into the diagonal part in the Hamiltonian of RMG. The $l=0, 1, 2, \dots$ is layer index of RMG. $d_0$ is the distance between the surface of substrate and the bottom graphene layer. $d_\text{g}=3.35$\,\AA \ is the intrinsic interlayer distance of the RMG. Solving the total Hamiltonian $H^0=H^{0}_\text{RMG}+H^0_\text{sub}$ that consists of the $\bk\cdot\p$ model of RMG and the non-interacting Hamiltonian of electron transferred to the substrate, we retrieve the layer-resolved excess electron density distribution. This quantum-mechanically computed layer-resolved excess electron density then serves as the source term in the classical Poisson equation, which is solved with a dielectric constant $\epsilon_s$ to update the screened electrostatic potential. Simultaneously, the Fock energy correction $\delta E_\mathrm{CBM}$, evaluated from the carrier density in substrate $n_\text{sub}$, is added to the on-site energy of substrate. By performing a self-consistent solution of the aforementioned screening process, the actual layer-resolved charge distribution and the corresponding relative electrostatic potential energy shift of graphene layers with respect to the surface of substrate at equilibrium can be obtained.

\begin{figure}[bth!]
%\begin{center}
    \includegraphics[width=5in]{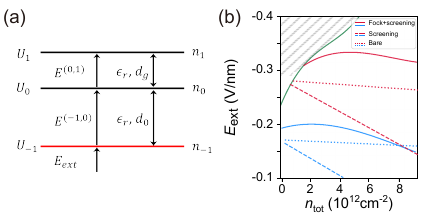}
\caption{(a) Schematic illustration of the self-consistent screening model of vertical electric field applied to bilayer graphene-based charge transfer heterostructure. $\epsilon_s$ is the dielectric constant. $d_0$ denotes the interlayer distance between the surface of substrate and the bottom graphene layer of RMG while $d_g$ denotes the intrinsic interlayer distance of the RMG. (b) Calculated phase boundaries for BLG-CrOCl under three levels of theoretical treatment: with initial band alignment denoted by ``Bare'' (dotted lines), with dielectric screening included denoted by ``Screening'' (dashed lines), and with both screening and Fock energy correction incorporated denoted by ``Screening+Fock'' (solid lines). The green, blue, and red curves denote, respectively, the phase boundaries separating (\i) non-charge-transfer and charge-transferred regions, (\ii) the hole-doped and charge neutrality regions, and (\iii) the charge neutrality and electron-doped regions.} 
%\end{center}
\label{SI_2}
\end{figure}

From Gauss' law, taking the bilayer graphene-substrate heterostructure as an example, the electric field $E^{(l,l+1)}$ between the $l$-th and $(l+1)$-th layers is
\begin{align}
    E^{(l, l+1)} - E_\text{ext} &= -\frac{n_l e}{\epsilon_0\epsilon_s} \quad \text{if} \quad l = -1\;, \nn
    E^{(l, l+1)} - E^{(l-1, l)} &= -\frac{n_l e}{\epsilon_0\epsilon_s} \quad \text{if} \quad l \ge 0\;,
\end{align}
for $l =-1, 0, 1$. The negative value of $l$ corresponds to the surface of the substrate while nonnegative $l$ correspond to RMG. $n_l$ is defined with respect to intralayer charge neutral point, i.e., positive if there is excess electron and vice versa. Then the layer-resolved on-site energies $U_l$ are set given that
\begin{align}
    U_{l} &= U_{l-1} + eE^{(l-1, l)}d_0 \quad \text{if} \quad l = 0\;, \nn
    U_{l} &= U_{l-1} + eE^{(l-1, l)}d_g \quad \text{if} \quad l \ge 1\;. \nn
\end{align}
We take the substrate surface as the reference point with zero electrostatic potential. 

% In our calculations, we take the limit $\bk_s \rightarrow \bk$, as we focus on low-energy states in both the RMG and substrate. In this regime, the distinction between wavevectors defined in the graphene and substrate Brillouin zones becomes negligible, since the relevant momenta lie deep within the respective Brillouin zones and are far from their boundaries. 

We have proposed that the characteristic features of RMG charge-transfer heterostructure, the pronounced ``broad'' and ``bent'' charge neutrality region of RMG, are attributed to the competition between the conventional screening effect and the Fock energy correction in the main text. To facilitate interpretation of this analysis, \figurename~\ref{SI_2}(b) illustrates the evolution of the upper (red) and lower (blue) boundary of the charge neutrality region across three levels of theoretical consideration. The nearly horizontal dotted lines (labeled ``Bare'') denote the boundaries of charge neutrality region obtained considering neither dielectric screening nor Fock energy correction. Inclusion of dielectric screening yields steeper linear dashed boundaries (labeled ``Screening''). Further incorporation of the Fock energy correction results in curved solid lines (labeled ``Screening$+$Fock''), which delineate the bent charge neutrality boundaries and enclose a markedly expanded charge neutrality region. Hence, our approach captures the essential physics governing interlayer charge redistribution and dielectric screening, thereby establishing a reliable foundation for understanding electron-electron interaction effects in graphene-based charge-transfer heterostructures.

\section{\BLUE{The dependence of phase boundaries on $d_0$ and $\gamma_m$}}
\label{sec:S4}
In the main text, we have validated our self-consistent electrostatic screening model using the BLG-CrOCl heterostructure, with the effective mass of the CBM of CrOCl $m^* \approx 1.3m_0$ and the energy offset with respect to the Dirac point of graphene $E_{\mathrm{CBM}} = -0.13$\,eV. The computed phase diagram for $d_0 = 4.5$\,\AA\,reproduces the experimentally observed charge neutrality region, whereas increasing $d_0$ to 8\,\AA\,dramatically shrinks it. As discussed in Sec.~\ref{sec:S2}, $\gamma_m$ affects the Fock self-energy correction to $E_{\mathrm{CBM}}$ and thus influences the charge-transfer mechanism. In this section, we investigate how the phase boundaries depend on these two parameters.

We first consider the band alignment conditions that determine the phase boundaries. For a fixed $n_{\mathrm{tot}}$,  the boundary point between the charge neutrality region and the electron-doped region corresponds to the alignment of the CBM of CrOCl with the CBM of BLG, while the boundary point between the hole-doped region and the charge neutrality region corresponds to the situation where the CBM of CrOCl aligns with the valence band maximum (VBM) of BLG. These alignment conditions can be expressed as follows:

\begin{align}
    E_{\mathrm{CBM}}+\delta E_{\mathrm{CBM}}(n_\mathrm{sub})+\epsilon_F(n_\mathrm{sub})&= eE_\mathrm{ext}^{(1)}d_0-\frac{n_\mathrm{sub}e^2} {\epsilon_0\epsilon_s}d_0 \\
    E_{\mathrm{CBM}}+\delta E_{\mathrm{CBM}}(n_\mathrm{sub})+\epsilon_F(n_\mathrm{sub})&= eE_\mathrm{ext}^{(2)}(d_0+d_g)-\frac{n_\mathrm{sub}e^2} {\epsilon_0\epsilon_s}(d_0+d_g) - \frac{n_\mathrm{bg}e^2}{\epsilon_0\epsilon_s}d_g 
\end{align}
At these boundary points, density of carriers transferred to the CrOCl satisfies $n_\mathrm{sub} = n_\mathrm{tot}$. The expression of $\delta E_{\mathrm{CBM}}(n_\mathrm{sub})$ id given by \ref{eq:fock_dE_aniso}, which is proportional to $\sqrt[4]{\gamma_m}K(\sqrt{1-\gamma_m})$. The Fermi energy $\epsilon_F(n_\mathrm{sub})=\frac{ n_\mathrm{sub}\pi\hbar^2}{2m^*}$ is a constant when the density-of state mass $m^*$ and $n_\mathrm{sub}$ are fixed. The $E_\mathrm{ext}^{(1)}$ and $E_\mathrm{ext}^{(12)}$ denote the external electric field at the upper boundary point and lower boundary point, respectively. And $n_{bg}$ denotes the layer-resolved carrier density of the bottom graphene sheet of BLG. 

\begin{figure}[bth!]
%\begin{center}
    \includegraphics[width=7in]{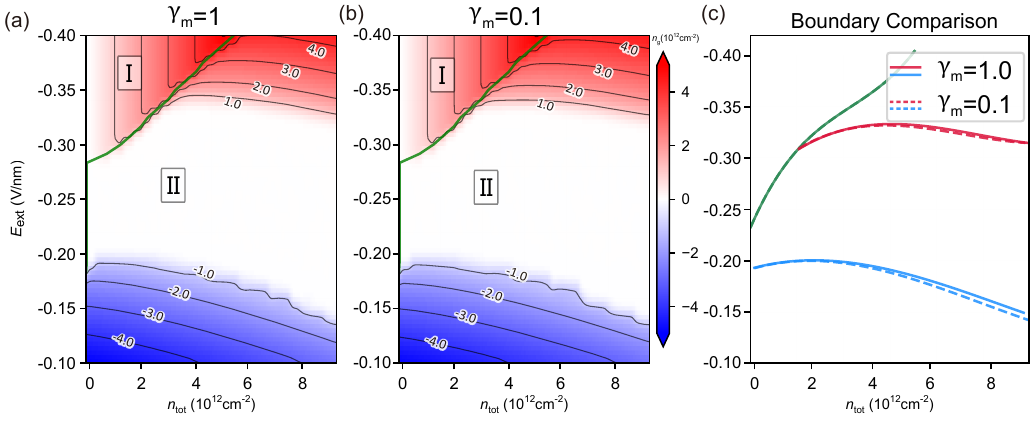}
\caption{(a) Carrier densities of the BLG $n_{\text{g}}$ in a BLG-CrOCl heterostructures with isotropic effective mass $m^*$. (b) Carrier densities of the BLG $n_{\text{g}}$ in a BLG-CrOCl heterostructures with anisotropic effective mass ($\gamma_m=0.1$). The interlayer distance is set to $d_0=4.5$\,\AA. The black solid lines mark iso-doping levels in BLG. (c) Comparison of the phase boundaries: solid lines correspond to the isotropic case (a), and dashed lines correspond to the anisotropic case (b).} 
%\end{center}
\label{SI_3}
\end{figure}

Then we analyze the dependence of phase boundaries on $d_0$ and $\gamma_m$. As mentioned in the main text, for small $n_\mathrm{sub}$ the Fock self‑energy correction dominates and bends the phase boundary upward, whereas at larger $n_\mathrm{sub}$ the dielectric screening effect takes over and its linear correction causes the boundary to decrease almost linearly. In this context, the mass anisotropy ratio 
$\gamma_m$ modulates the Fock correction, which becomes weaker when $\gamma_m$ deviates from the isotropic limit ($\gamma_m=1$). However, for $\gamma_m=0.1$, a strong deviation from isotropy, the Fock correction drops by only about 10\%. As shown in \figurename~\ref{SI_3}(a) and (b), the phase diagrams of carrier densities $n_{\text{g}}$ in a BLG-CrOCl heterostructures for the isotropic model $\gamma_m=1$ and the anisotropic model $\gamma_m=0.1$ appear nearly identical. We further compare the phase boundaries obtained from the anisotropic model (dashed lines) and the isotropic model (solid lines) in \figurename~\ref{SI_3}(c). Although the mass anisotropy ratio $\gamma_m=0.1$ reduces the Fock correction and thereby shifts the transition from Fock to screening dominance to smaller $n_\mathrm{sub}$, which manifests as a slight leftward movement of the linearly descending part of the phase boundaries, the overall change is minimal and can be safely neglected. We therefore adopt the isotropic approximation in our calculations.

\begin{figure}[bth!]
%\begin{center}
    \includegraphics[width=7in]{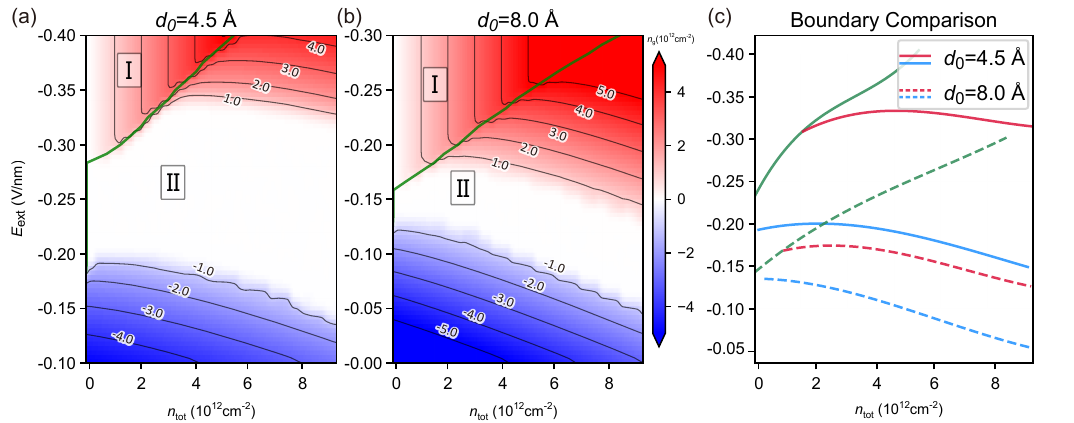}
\caption{(a) Carrier densities of the BLG $n_{\text{g}}$ in a BLG-CrOCl heterostructures with interlayer distance $d_0=4.5$\,\AA. (b) Carrier densities of the BLG $n_{\text{g}}$ in a BLG-CrOCl heterostructures with interlayer distance $d_0=8$\,\AA. The black solid lines mark iso-doping levels in BLG. (c) Comparison of the phase boundaries: solid lines correspond to $d_0=4.5$\,\AA (a), and dashed lines correspond to $d_0=8$\,\AA (b).
} 
%\end{center}
\label{SI_4}
\end{figure}

Under this approximation, we then examine the role of the interlayer distance $d_0$: increasing the interlayer distance $d_0$ enhances the dielectric screening, so that the crossover from the Fock‑dominated to the screening‑dominated regime occurs at a smaller $n_\mathrm{sub}$ as well. In addition, a larger $d_0$ is accompanied by a smaller external field $|E_\mathrm{ext}^{(1)}|$, $|E_\mathrm{ext}^{(2)}|$ and their difference $\delta E_\mathrm{ext} \propto 1/d_0(d_0+d_g)$. As a result, the charge neutrality region shifts downward and significantly shrinks in size with increasing $d_0$. At the same time, the crossover point concurrently moves to lower $n_\mathrm{sub}$. In \figurename~\ref{SI_4}(a) and (b), we present the phase diagrams of the carrier densities of the BLG $n_{\text{g}}$ in a BLG-CrOCl heterostructures for $d_0=4.5$\,\AA and $d_0=8$\,\AA, respectively. And \figurename~\ref{SI_3}(c) shows the comparison of the phase boundaries for them, which consists of our analysis. 

In our calculations, the $\bk$ mesh is $600 \times 600$ and we use the dielectric constant $\epsilon_s=14$ and $\epsilon_\text{Cr} =10 $ for the BLG-CrOCl heterostructures.

\section{Continuum model describing rhombohedral multilayer graphene coupled to insulating substrate via interlayer Coulomb coupling}
\label{sec: Continuum model describing rhombohedral multilayer graphene coupled to insulating substrate via interlayer Coulomb coupling}
Now we discuss the interlayer coupling between electrons within RMG and electrons transferred to substrate. Since the interlayer hopping amplitude decays exponentially with interlayer separation $d_0$ while the interlayer Coulomb interaction decays only as a power law, the former can be safely neglected when the interlayer separation $d_0$ and the lattice mismatch is large. Hence, the low energy electrons in this system have an emergent approximate layer U(1) symmetry. The interlayer coupling is dominated by the long-range Coulomb potential between electrons in the substrate and those in the RMG. In this regime, we can treat such Coulomb-coupled two-component systems in the ``Born-Oppenheimer'' type approximation, where the wavefunction of the ground state can be written as the product of that of RMG and that of substrate, i.e.,
\begin{equation}
    \left|\bm{\Psi}\right\rangle=\left|\bm{\Psi}\right\rangle_c \otimes \left|\bm{\Psi}\right\rangle_d,
\end{equation}
where the ground wavefunction with low index $c$ and $d$ represents that of RMG and substrate, respectively. As discussed in the main text, given that the carrier density of the substrate  $n_\text{sub}$ is below a threshold and the effective mass $m^*$ at CBM is large, long-wavelength charge order state can be induced at the surface of the substrate via Coulomb interactions, with electrons localized into a superlattice. The geometry of superlattice is determined by the long-wavelength order at the interface. In our calculation, we set the superlattice to be triangular. The Hamiltonian between RMG and the insulating substrate reads:
\begin{equation}
   \begin{aligned}
       \hat{H}_\text{G-S}
       = \int{d^2\br d^2\br'\sum_{\sigma \sigma'}{\hat{\psi}^{\dagger}_{c,\sigma} (\br)\hat{\psi}^{\dagger}_{d,\sigma'}(\br')
       V|\br-\br'| \hat{\psi}_{d,\sigma'}(\br') \hat{\psi}_{c,\sigma}(\br)}},  
   \end{aligned}
\end{equation}
where the field operators $\hat{\psi}(\br)$ with subscript $c$ and $d$ act on electrons in RMG and substrate, respectively. $V(\mathbf{r} - \mathbf{r}') = e^2/(4\pi \epsilon_0 \epsilon_r |\mathbf{r} - \mathbf{r}'|)$ represents the three-dimensional Coulomb potential, with $\epsilon_0$ and $\epsilon_r$ denoting the vacuum permittivity and the relative dielectric constant of the environment, respectively. $l$ is the layer index. $\sigma$ is spin index.

In terms of Wannier functions $\phi_{l\alpha}$ localized on the atomic lattice sites $\bm{a}_i$ of $l$th graphene layer within RMG and $\tilde{\phi}$ localized on superlattice sites $\bm{R}_i$  formed on the surface of substrate, the field operators can be represented as:
\begin{equation}
    \begin{aligned}
        \hat{\psi}^{\dagger}_{c,\sigma}(\br) &= \sum_{i,l\alpha}{\phi^*_{l\alpha}(\br - \bm{a}_i-\bm{\tau}_{\alpha} -d_l\hat{\bm{z}}) \chi_\sigma^{\dagger} \hat{c}^\dagger_{i l\alpha, \sigma}} \\
        \hat{\psi}^{\dagger}_{d,\sigma}(\br) &= \sum_{i} {\tilde{\phi}^*(\br-\bm{R}_i)
        \chi_\sigma^{\dagger} \hat{d}^\dagger_{i, \sigma}}
    \end{aligned}.
\end{equation}
Here $i, l, \alpha, \sigma$ represent lattice site index, layer index, sublattice index and spin index, respectively. $\chi_\sigma$ is two component spinor wavefunction. It is worthwhile to note that here the Bravais lattice sites and the corresponding Wannier functions for the substrate refer to those of the spontaneously generated charge ordered superlattice, not the atomic lattices of the substrate. After transforming the Hamiltonian $\hat{H}_\text{G-S}$ into the Wannier basis, we obtain:
\begin{equation}
    \hat{H}_\text{G-S} = \sum_{\substack{\sigma,\sigma'\\\alpha, \alpha'\\l,l'}}\sum_{\substack{i,i'\\j,j'}} U^{\sigma,\sigma'}_{il\alpha j, i' l'\alpha'j'}\hat{c}^\dagger_{il\alpha,\sigma}\hat{d}^\dagger_{j,\sigma'}\hat{d}_{j'\sigma'}\hat{c}_{i'l'\alpha',\sigma}
\end{equation}
with
\begin{equation}
    U^{\sigma,\sigma'}_{il\alpha j, i' l'\alpha'j'}=\int{d^2\br d^2\br'} \phi^*_{l\alpha}(\br - \bm{a}_i-\bm{\tau}_{\alpha} -d_l\hat{\bm{z}})\tilde{\phi}^*(\br-\bm{R}_j)\tilde{\phi}(\br-\bm{R}_{j'})\phi_{l'\alpha'}(\br - \bm{a}_{i'}-\bm{\tau}_{\alpha'} -d_{l'}\hat{\bm{z}})\chi^\dagger_\sigma\chi^\dagger_{\sigma'}\chi_{\sigma'}\chi_{\sigma}. 
\end{equation}
We assume that the Wannier functions are so localized that
\begin{align}
    \phi^*_{l\alpha}(\br - \bm{a}_i-\bm{\tau}_{\alpha} -d_l\hat{\bm{z}})\phi_{l'\alpha'}(\br - \bm{a}_{i'}-\bm{\tau}_{\alpha'} -d_{l'}\hat{\bm{z}}) & \approx 0 \quad \text{if} \quad (i,l, \alpha) \ne (i',l',\alpha') \\
    \tilde{\phi}^*(\br-\bm{R}_j)\tilde{\phi}(\br-\bm{R}_j') & \approx 0 \quad \text{if} \quad j \ne j' \\
    \left |  \phi_{l\alpha}(\br - \bm{a}_i-\bm{\tau}_{\alpha} -d_l\hat{\bm{z}}) \right |^2 &\approx \delta^2 (\br - \bm{a}_i-\bm{\tau}_{\alpha} -d_l\hat{\bm{z}}) \\
    \left | \tilde{\phi}(\br-\bm{R}_j) \right |^2 & \approx \delta^2(\br-\bm{R}_j),  
\end{align}
where $\delta^2(\br)$ is 2D Dirac $\delta$-function distribution. Then, the expression of $U^{\sigma,\sigma'}_{il\alpha j, i' l'\alpha'j'}$ can be simplified to $U_{il\alpha j}\delta_{i,i'}\delta_{l,l'}\delta_{\alpha,\alpha'}\delta_{j,j'}$ with the Kronecker delta $\delta_{\mu,\nu}$ and $U_{il\alpha,j} = V(|\bm{a}_i+\bm{\tau}_{\alpha} +d_l\hat{\bm{z}}- \R_j|)$. 

Next, using the Fourier transformation
\begin{equation}
    \begin{aligned}
         \hat{c}_{il\alpha, \sigma} &= \frac{1}{N_c}\sum_{\bk} e^{i\bk\cdot\bm{a}_i}\hat{c}_{l\alpha,\sigma}(\bk) \\
         \hat{d}_{i,\sigma} &= \frac{1}{N_d}\sum_{\tilde{\bk}} e^{i\tilde{\bk}\cdot\R_i}\hat{d}_{\sigma}(\tilde{\bk})
    \end{aligned}
\end{equation}
where $N_c$ and $N_d$ are the number of lattice sites for electron in RMG and the substrate, respectively. $\bk$ and $\tilde{\bk}$ are the wavevectors in the Brillouin zone of graphene and that of the long-wavelength superlattice in the substrate, respectively. Then we can write the Hamiltonian $\hat{H}_\text{G-S}$ in reciprocal space:
\begin{equation}
    \hat{H}_\text{G-S} = \frac{1}{N_c N_d}\sum_{\substack{\bk,\bk'\\ \tilde{\bk},\tilde{\bk'}}}\sum_{\substack{\sigma,\sigma'\\ il\alpha j}} U_{il\alpha j}e^{i(\bk'-\bk)\cdot(\bm{a}_i-\R_j)}e^{i(\bk'-\bk+\tilde{\bk'}-\tilde{\bk})\cdot\R_j} \hat{c}^\dagger_{l\alpha,\sigma}(\bk)\hat{d}^\dagger_{\sigma'}(\tilde{\bk})\hat{d}_{\sigma'}(\tilde{\bk'})\hat{c}_{l'\alpha',\sigma}(\bk').
\end{equation}
We define $\tilde{\R} = \bm{a}_i-\R_j$ and set $\bk-\bk'=\q=\tilde{\q}+\G$, where $\tilde{q}$ is wave vector defined within the mini-Brillouin zone of the superlattice and $\G$ is a reciprocal vector of the superlattice. Utilizing the identity $\sum_j e^{i(\tilde{\bk}'-\tilde{\bk}-\tilde{\q}-\G)\cdot\tilde{\R}_j} = N_d\delta_{\tilde{\bk}'-\tilde{\bk},\tilde{\q}}$, the above expression is simplified as:
\begin{equation}
    \hat{H}_\text{G-S} = \sum_{\sigma\sigma',l\alpha}\sum_{\substack{\bk,\tilde{\bk} \\ \tilde{\q}, \G}} \tilde{V}(\tilde{\q}+\G) \hat{c}^\dagger_{\sigma,l\alpha}(\bk)\hat{d}^\dagger_{\sigma'}(\tilde{\bk})\hat{d}_{\sigma'}(\tilde{\bk}+\tilde{\q})\hat{c}_{\sigma,l\alpha}(\bk-\tilde{\q}-\G).
\end{equation}
The expression of interlayer coupling $\tilde{V}(\tilde{\q}+\G)$ is:
\begin{equation}
    \begin{aligned}
        \tilde{V}(\tilde{\q}+\G) &= \frac{1}{N_c}\sum_i V(|\bm{a}_i+\bm{\tau}_\alpha+d_l\hat{\bm{z}}-\R_j|) e^{-i(\tilde{\q}+\G)\cdot(\bm{a}_i-\R_j)} \\
        & = \frac{1}{N_c}\sum_{\tilde{\R}} V(|\bm{\tau}_\alpha+d_l\hat{\bm{z}}+\tilde{\R}|) e^{-i(\tilde{\q}+\G)\cdot\tilde{\R}} \\
        & = \frac{1}{N_d \Omega_d}\int d^2 \br V(|\br +\bm{\tau}_\alpha+d_l\hat{\bm{z}}|) e^{-i(\tilde{\q}+\G)\cdot\br} \\
        & = \frac{e^2}{2\epsilon_0\epsilon_rN_d\Omega_d} \frac{e^{-|\tilde{\q}+\G|d_l}}{|\tilde{\q}+\G|}    
    \end{aligned}
\end{equation}
where $\Omega_d$ is the area of the superlattice unit cell.  In the third line of the above derivation, we smear the sum over $\tilde{\R}$ by replacing it with an integral over the surface $S=N_d\Omega_d=N_c\Omega_c $ with $\Omega_c$ representing the area of graphene’s unit-cell since we are interested in the physics in the length scale of the superlattice {$\R_j$}, which is supposed to much larger than that of graphene. Finally, the last line is the 2D partial Fourier transformation of the 3D Coulomb potential. 

Since we concentrate on low energy physics, the operators of the electrons within RMG are introduced a valley index $\mu$. We neglected the intervalley interactions because of the exponential decay of $\tilde{V}(\tilde{\bq}+\G)$ and rewrite the Hamiltonian $\hat{H}_\text{G-S}$ as.\:
\begin{equation}
     \hat{H}_\text{G-S} = \sum_{\substack{\sigma\sigma' \\ \mu l\alpha }} \sum_{\substack{\bk,\tilde{\bk} \\ \tilde{\bq},\G}}
    \tilde{V}(\tilde{\bq}+\G) \hat{c}^\dagger_{\sigma l\mu\alpha}(\bk) \hat{d}^\dagger_{\sigma'}(\tilde{\bk})
    \hat{d}_{\sigma'}(\tilde{\bk}+\tilde{\bq}) \hat{c}_{\sigma l\mu\alpha}(\bk - \tilde{\bq}-\G).
\end{equation}

In the Hartree approximation by contracting fermion operators with subscript $c$ and $d$ separately, we obtain:
\begin{equation}
    \begin{aligned}
        \hat{H}_\text{G-S} &=  \sum_{\sigma\mu,l\alpha}\sum_{\bk,\G} \tilde{V(\G)}\sum_{\sigma',\tilde{\bk}'} \left \langle \hat{d}^\dagger_{\sigma'}(\tilde{\bk})\hat{d}_{\sigma'}(\tilde{\bk})\right\rangle \hat{c}^\dagger_{\sigma\mu l\alpha}(\bk)\hat{c}_{\sigma\mu l\alpha}(\bk-\G)  \\
        &= \nu_\mathrm{s} N_d \sum_{\sigma\mu,l\alpha}\sum_{\bk,\G} \tilde{V}(\G)\hat{c}^\dagger_{\sigma \mu\alpha}(\bk)\hat{c}_{\sigma \mu\alpha}(\bk-\G)
    \end{aligned}
\end{equation}
where $\nu_\mathrm{s}$ denotes the number of electrons within a superlattice unit cell. Writing $\bk = \tilde{\bk}+\G$ with $\G$ in the superlattice reciprocal lattice,the form of the coupling between RMG and insulating substrate reads:
\begin{equation}
    \hat{H}_\text{G-S} = \sum_{\sigma\mu l\alpha}\sum_{\G,\Q} \tilde{U}_\text{s}(\Q) \hat{c}^{\dagger}_{\sigma\mu l\alpha,\G+\Q}(\tilde{\bk})\hat{c}_{\sigma\mu l\alpha,\G}(\tilde{\bk}),    
\end{equation}

where $\tilde{U}_\text{s}(\Q) = \frac{\nu_\text{s} e^2}{2\epsilon_0\epsilon_r\Omega_d}\frac{e^{-|\Q|d}}{|\Q|}$.

Incorporating the superlattice potential, we obtain the low-energy effective continuum model for the RMG part:
\begin{equation}
    \hat{H}_{\text{RMG}}^{0, \mu} = \hat{H}_\text{RMG}^{0, \mu} + U_\text{s}(\br) + U_\text{d}
\end{equation}
where $U_\text{d}$ is the electrostatic on-site energy obtained from self-consistent screening model. $U_\text{s}(\br)$ is the background superlattice potential with the period $U_\text{s}(\br) = U_\text{s}(\br+\bf{L_s})$, whose Fourier components are exactly $\tilde{U}_\text{s}(\Q)$. Such a continuum-model description is adopted throughout the calculations within ``Born-Oppenheimer-type" treatment. Given that $L_s \ge a$ ( $a=2.46$\,\AA \ is the lattice constant of graphene) is always fulfilled with low carrier density of the substrate $n_\mathrm{sub}$.

\section{Renormalization of the model parameters due to interactions with remote-band electrons}
\label{sec: Renormalization of the model parameters due to interactions with remote-band electrons}
In our work, we mostly focus on the low-energy physics around charge neutrality point. We set up a low-energy window marked by $E_C^{*}$ which roughly includes three conduction subbands and three valence subbands per spin per valley. Electron-electron interactions  become important and non-perturbative within this low-energy window within $E_C^*$, while they can be approximately treated as perturbations to single-particle energy outside $E_C^*$. We call the electrons within the low-energy window as low-energy electrons, while those out-side $E_C^*$ as remote-band electrons. Although we focus on the low-energy physics, the occupied remote-band electrons indeed play an important role considering $e$-$e$ interactions. The electrons in the filled remote bands will act through long-ranged Coulomb potential upon the properties of low-energy electrons. As a result, an effective low-energy single-particle Hamiltonian  (within $E_C^*$) would have parameters in general larger in amplitudes than the non-interacting ones. For example, it is well known that the Fermi velocity around the Dirac point in graphene would be amplified by the filled Dirac Fermi sea \cite{elias_natphys2011}. We take into account this effect using perturbative renormalization group (RG) approach \cite{vafek_prl2020,guozhq-fci-prb2024}. Without going into detailed derivations, we directly write down the expressions of the renormalized model parameters as already reported in Ref.~\cite{guozhq-fci-prb2024}
\begin{subequations}
\begin{align}
 v_F(E_C^*)&=v_F^0 \left(1+\frac{\alpha_0}{4\epsilon_r}\log{\frac{E_C}{E_C^*}} \right)\; \label{eq:H-RG-a}\\
 t_{\perp}(E_C^*)&=t_{\perp}\,\left(1+\frac{\alpha_0}{4\epsilon_r}\log{\frac{E_C}{E_C^*}}\right)\; \label{eq:H-RG-c}\\
 v_{\perp}(E_C^*)&=v_{\perp}\;.
 \label{eq:H-RG-f}
\end{align}
\label{eq:H-RG}
\end{subequations}
Here $E_C$ is the largest energy cutoff of the continuum model above which the Dirac-fermion description to graphene would no longer be valid. $\alpha_0=e^2/4\pi\epsilon_0\hbar v_F$ is the effective fine structure constant of graphene. We refer the readers to Ref.\cite{guozhq-fci-prb2024, lu-nc23} for detailed derivations of the above equations.

Using the aforementioned low-energy effective continuum model for RMG with parameters corrected by RG approach, we have calculated the single-particle properties for bilayer (BLG)  on a CrOCl substrate. In the main text, isolated flat bands emergent in the energy band structure of BLG with parameters $E_\text{ext} = -0.225$\, V/nm and $n_\text{tot} = 4.5 \times 10^{11}$\,cm$^{-2}$ in the presence of superlattice potential $U_s$, as illustrated in Fig.~1(b) of the main text.   

\begin{figure}[bth!]
%\begin{center}
    \includegraphics[width=5in]{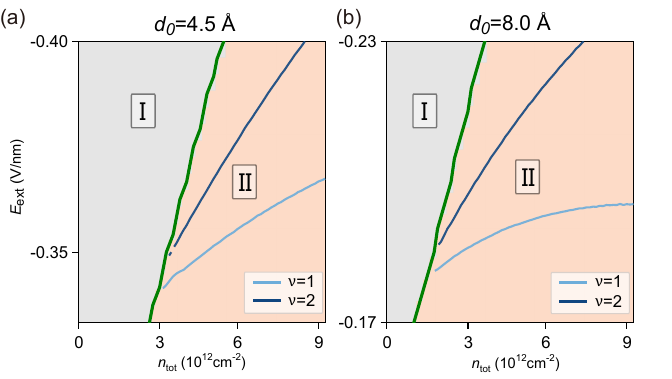}
\caption{Phase diagrams of the Chern number for the LCB of BLG-CrOCl with interlayer distance $d_0=4.5$\,\AA (a) and $d_0=8$\,\AA (b), computed using the effective continuum model of RMG with parameters renormalized via the RG approach. The green solid line represent the phase boundary between no charge transfer region and charge transfer region. The non-zero Chern number $|C|=1$ region is marked by light orange.} %\sout{The shaded regions correspond to parameters where the lowest conduction band (LCB) has a direct gap separating it with high-energy bands  $\lessapprox 1 $\,meV, precluding a well-defined Chern number in numerical evaluation due to finite mesh. Along iso-filing lines with filling factors $\nu=1$ and $\nu=2$, however, the LCB is well isolated, enabling unambiguous evaluation of its topological invariant.}}
\label{SI_5}
%\end{center}
\end{figure}

\BLUE{We have calculated the Chern number of the lowest conduction band (LCB) of the non-interacting superlattice bands from K valley of graphene. In our model, graphene coupled with a superlattice potential exhibits a valley U(1) symmetry, since the intervalley scattering from the superlattice potential are negligibly small for the superlattice constant much larger than graphene’s atomic lattice constant considered. This valley U(1) symmetry allows us to define the Chern numbers of the subbands in each valley, and they possess opposite signs due to time-reversal symmetry. For the non-interacting superlattice bands, we obtain the Chern number via numerically integrating the Berry curvature over the superlattice mini Brillouin zone:
\begin{equation}
    C=\frac{1}{2\pi}\int_{\text{BZ}}d^2\k \,{\Omega(\k)},
\end{equation}
where $\Omega(\k) = \nabla_{\k}\times\left\langle u(\k)\right|i\nabla_{\k}\left|u(\k)\right\rangle$. This formula is numerically implemented using the widely adopted method reported in Ref.\cite{Chern-number-2005}. This method is gauge independent, and no need to ensure gauge continuity throughout the mini Brillouin zone. In \figurename~\ref{SI_5}, we present phase diagrams of the Chern number of lowest conduction band (LCB) of  BLG-CrOCl heterostructure  with interlayer distance $d_0=4.5$\,\AA (a) and  $d_0=8$\,\AA (b), both of which display a nonzero Chern number $|C|=1$ across nearly the entire charge-transfer region. These nontrivial topological flat bands amplify interactions effect and potentially promote the formation of topological correlated states.}

\begin{figure}[bth!]
%\begin{center}
    \includegraphics[width=5in]{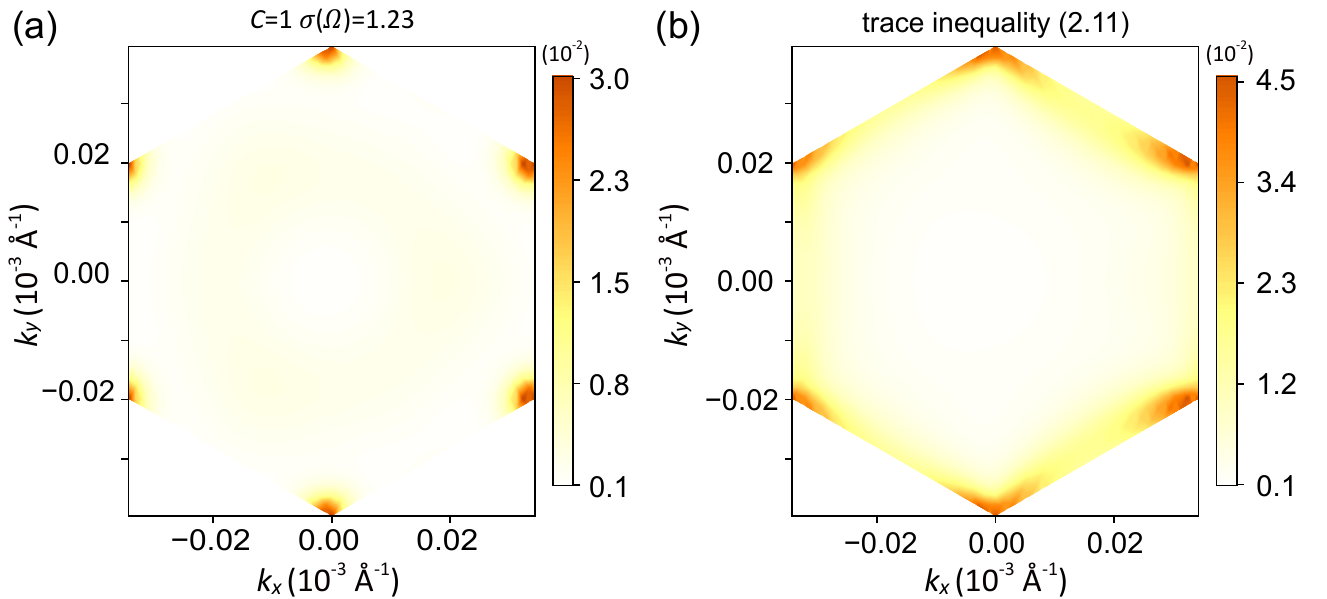}
\caption{The distribution of (a) berry curvature and (b) the normalized local geometric deviation from the ideal trace condition $(\text{Tr}(g)-|\Omega|)⁄2\pi$ of the lowest conduction band of the BLG-CrOCl heterostructure with parameters $d_0=4.5$\,\AA, $E_\text{ext}=-0.3417 $\,$\text{V}/\text{nm}$ and $n_\text{tot}= 3.14\times10^{12}$\,cm$^{-2}$, computed using the effective continuum model of RMG with parameters renormalized via the RG approach. The Chern number is $C=1$ and the normalized berry curvature standard deviation $\sigma(\Omega)=1.23$. The difference between the integral of trace of Fubini-Study metric $\text{Tr}(g)$ of the Bloch states and Chern number $|C|$ is 2.11.}
\label{SI_6}
%\end{center}
\end{figure}

\BLUE{To characterize the quantum geometric properties of the lowest conduction band, we further compute the distributions of the berry curvature and $(\text{Tr}(g)-|\Omega(\k)|)/2\pi$ over the entire Brillouin zone of the lowest conduction band of the BLG-CrOCl heterostructure with parameters $d_0=4.5$\,\AA, $E_\text{ext}=-0.3417 $\,$\text{V}/\text{nm}$ and $n_\text{tot}= 3.14\times10^{12}$\,cm$^{-2}$, as shown in \figurename~\ref{SI_6}, which are the key quantities that determine the feasibility of FCI states. Our analysis shows that the ideal conditions required for FCI states are significantly violated. Using exact diagonalization, we therefore have not identified FCI states in the BLG-CrOCl system considered in this work. Nevertheless, different substrate materials, heterostructure geometries, dielectric environments, or superlattice periods may lead to topological flat bands with more uniform Berry curvature and more ideal quantum geometry, thereby providing more favorable conditions for FCI phases.}

\section{Band-projected Hartree-Fock method}
\label{sec: Band-projected Hartree-Fock method}
In this section, we provide the derivation of the e-e interaction Hamiltonian in the Hartree-Fock (HF) approximation projected to the band basis.

We consider the $e$-$e$ Coulomb interactions
\begin{equation}
\hat{V}_\text{ee}=\frac{1}{2}\int d^2 r  d^2 r' \sum _{\sigma, \sigma '} \hat{\psi}_\sigma ^{\dagger}(\br)\hat{\psi}_{\sigma '}^{\dagger}(\br ') V_c (|\br -\br '|) \hat{\psi}_{\sigma '}(\br ') \hat{\psi}_{\sigma}(\br)
\label{eq:coulomb}
\end{equation}
where $\hat{\psi}_{\sigma}(\br)$ is real-space electron annihilation operator at $\br$ with spin $\sigma$. Such interaction can be expressed in Wannier basis as 
\begin{equation}
\hat{V}_\text{ee}=\frac{1}{2}\sum _{i i' j j'}\sum_{l_1 l_1' l_2 l_2'}\sum _{\alpha \alpha '\beta \beta ' }\sum _{\sigma \sigma '} \hat{c}^{\dagger}_{i, \sigma l_1 \alpha}\hat{c}^{\dagger}_{i', \sigma ' l_1' \alpha '} V^{\alpha \beta l_1 l_2 \sigma , \alpha ' \beta ' l_1' l_2' \sigma '} _{ij,i'j'}\hat{c}_{j', \sigma ' l_2' \beta '} \hat{c}_{j, \sigma l_2 \beta}\;,
\end{equation}
where
\begin{align}
&V^{\alpha \beta l_1 l_2 \sigma , \alpha ' \beta ' l_1' l_2' \sigma '} _{ij,i'j'} \nonumber \\
&=\int d^2 r d^2 r'  V_c (|\br -\br '|) \,\phi ^*_{l_1,\alpha} (\mathbf{r}-\mathbf{R}_i-\bm{\tau}_{l_1,\alpha})\,\phi_{l_2,\beta} (\mathbf{r}-\mathbf{R}_j- \bm{\tau}_{l_2,\beta}) \phi^*_{l_1', \alpha  '}(\br-\mathbf{R}_i'-\bm{\tau}_{l_1' , \alpha '})\phi _{l_2', \beta  '}(\br-\mathbf{R}_j'-\bm{\tau} _{l_2', \beta '}) \nonumber \\
&\quad \times \chi ^{\dagger}_\sigma \chi ^{\dagger}_{\sigma '}\chi _{\sigma '}\chi _{\sigma} .
\end{align}
Here $i$, $\alpha$, $l_1$($l_2$), and $\sigma$ refer to atomic lattice vectors, sublattice index, layer index and spin index, respectively. $\phi$ is Wannier function and $\chi$ is the two-component spinor wave function. We further assume that the ``density-density" like interaction is dominant in the system, i.e., $V^{\alpha \beta l_1 l_2 \sigma , \alpha ' \beta ' l_1' l_2' \sigma'} _{ij,i'j'}\approx V^{\alpha \alpha l_1 l_1 \sigma , \alpha ' \alpha ' l_1' l_1' \sigma '} _{ii,i'i'}\equiv V_{i \sigma l \alpha  ,i' \sigma ' l' \alpha '}$, then the Coulomb interaction is simplified 
\begin{align}
\hat{V}_\text{ee}=&\frac{1}{2}\sum _{i i'}\sum _{\alpha \alpha ', l l'}\sum _{\sigma \sigma '}\hat{c}^{\dagger}_{i, \sigma l \alpha}\hat{c}^{\dagger}_{i', \sigma' l' \alpha} V_{i\sigma l \alpha, i' \sigma ' l' \alpha '}\hat{c}_{i', \sigma ' l' \alpha '}\hat{c}_{i, \sigma l \alpha} \nonumber \\
\approx&\frac{1}{2}\sum _{i l \alpha \neq i' l' \alpha '}\sum _{\sigma \sigma '}\hat{c}^{\dagger}_{i, \sigma l \alpha} \hat{c}^{\dagger}_{i', \sigma ' l' \alpha '} V_{i l \alpha,i' l' \alpha '}\hat{c}_{i', \sigma ' l' \alpha '}\hat{c}_{i, \sigma l \alpha} \nonumber 
\end{align}
Here we neglect atomistic on-site Coulomb interactions which is at least one order of magnitude weaker than long-range inter-site Coulomb interactions in the context of moir\'e superlattice and other long-period superlattices \cite{zhang_prl2022}. Given that the electron density is low ($10^{12}$~cm$^{-2}$ in our problem),  the chance that two electrons meet at the same atomic site is very low. Therefore, the Coulomb interactions between two electrons are mostly contributed by the inter-site ones. 

In order to model the screening effects  and capture the layer dependence of Coulomb interactions in multilayer graphene, we introduce the following Coulomb potential in momentum space
\begin{align}
    &V_{ll} (\mathbf{q})=\frac{e^2}{2 \Omega_d \epsilon_r \epsilon_0 \sqrt{q^2+\kappa^2}}\;\nn
    &V_{ll'}(\mathbf{q})=\frac{e^2}{2 \Omega_d \epsilon_r \epsilon_0 q} e^{-q|l-l'|d_0},\,\hspace{6pt}\, \text{if } l\neq l'
    \label{eq:Vq}
\end{align}
where $\Omega _d$ is the area of the superlattice's primitive cell and $\kappa = 1/400$\,\AA$^{-1}$ is the inverse screening length. 

Since we are interested in the low-energy bands, the intersite Coulomb interactions can be divided into the intra-valley term and the inter-valley term. The intervalley term is at least two orders of magnitudes weaker than the intravalley one in our system because of the small Brillouin zone of the superlattice, thus is neglected in our present study. The intra-valley term $\hat{V}^{\text{intra}}$ is expressed as
\begin{equation}
\hat{V}^{\rm{intra}}=\frac{1}{2N_s}\sum_{\alpha\alpha ', l l'}\sum_{\mu\mu ',\sigma\sigma '}\sum_{\bk \bk ' \bq} V_{l l'}(\bq)\,
\hat{c}^{\dagger}_{\sigma \mu l \alpha}(\bk+\bq) \hat{c}^{\dagger}_{\sigma' \mu ' l' \alpha '}(\bk ' - \bq) \hat{c}_{\sigma ' \mu ' l' \alpha '}(\bk ')\hat{c}_{\sigma \mu l \alpha}(\bk)\;,
\label{eq:h-intra}
\end{equation}
where $N_s$ is the total number of the superlattice's sites. 

The electron annihilation operator can be transformed from the original basis to the band basis:
\begin{equation}
\hat{c}_{\sigma\mu l \alpha}(\bk)\equiv \hat{c}_{\sigma\mu l \alpha \G}(\btk) =\sum_n C_{ \mu l \alpha \mathbf{G},n}(\btk)\,\hat{c}_{\sigma \mu,n}(\btk)\;,
\label{eq:transform}
\end{equation}
where $C_{\mu l \alpha \mathbf{G},n}(\btk)$ is the expansion coefficient in the $n$-th Bloch eigenstate at $\btk$ of valley $\mu$: 
\begin{equation}
\ket{\sigma \mu, n; \btk}=\sum_{l \alpha \mathbf{G}}C_{\mu l \alpha \mathbf{G},n}(\btk)\,\ket{ \sigma, \mu, l, \alpha, \mathbf{G}; \btk }\;.
\end{equation}
We note that the non-interacting Bloch functions are spin degenerate. Using the transformation given in Eq.~(\ref{eq:transform}), the intra-valley Coulomb interaction can be written in the band basis
\begin{align}
\hat{V}^{\rm{intra}}&=\frac{1}{2N_s}\sum _{\btk \btk'\btq}\sum_{\substack{\mu\mu' \\ \sigma\sigma'\\l l'}}\sum_{\substack{nm\\ n'm'}} \left(\sum _{\mathbf{Q}}\,V_{ll'} (\mathbf{Q}+\btq)\,\Omega^{\mu l ,\mu' l'}_{nm,n'm'}(\btk,\btk',\btq,\mathbf{Q})\right) \nonumber \\
&\times \hat{c}^{\dagger}_{\sigma\mu,n}(\btk+\btq) \hat{c}^{\dagger}_{\sigma'\mu',n'}(\btk'-\btq) \hat{c}_{\sigma'\mu',m'}(\btk') \hat{c}_{\sigma\mu,m}(\btk)
\label{eq:Hintra-band}
\end{align}
where the form factor $\Omega ^{\mu  l,\mu' l'}_{nm,n'm'}$ are written respectively as
\begin{equation}
\Omega ^{\mu l ,\mu' l'}_{nm,n'm'}(\btk,\btk',\btq,\mathbf{Q})
=\sum _{\alpha\alpha'\mathbf{G}\mathbf{G}'}C^*_{\mu l \alpha\mathbf{G}+\mathbf{Q},n}(\btk+\btq) C^*_{\mu'l'\alpha'\mathbf{G}'-\mathbf{Q},n'}(\btk'-\btq)C_{\mu'l'\alpha'\mathbf{G}',m'}(\btk')C_{\mu l \alpha\mathbf{G},m}(\btk).
\end{equation}

In our calculations, the long-range Coulomb interactions are projected onto the wavefunctions of three conduction bands and three valence bands (per valley per spin) obtained from the RG-corrected low-energy continuum model. Then we carry out unrestricted HF calculations within this 24-band low-energy subspace (including spin and valley degrees of freedom). To explore possible symmetry-broken phases, we consider 32 distinct trial initial states, each characterized by order parameters of the form {$s^{0, z}\tau^a\sigma^b$} with $a,b = 0, x, y,z$, where $\bm{s}$, $\bm{\tau}$ and $\bm{\sigma}$ denote Pauli matrices acting in spin, valley, and sublattice spaces, respectively. %\sout{The dielectric constant $\epsilon_r =8$ is for BLG and $\epsilon_r = 16$ is for TriLG}. 

\BLUE{In Fig.2(a) and (b) of the main text, we have shown the single-particle spectra of Hartree-Fock ground states of BLG-CrOCl heterostructure with $d_0=4.5$\,\AA\, at integer filling factors $\nu=1$ and $\nu=2$, respectively. At filling factor $\nu=1$, the ground state is a spin-valley polarized Chern insulator with $|C|=1$, characterized by the order parameters $\left\langle s_z\right\rangle=\left\langle\tau_z\right\rangle=\left\langle s_z\tau_z\right\rangle=1 $. These equalities indicate full polarization of both spin and valley flavors. Consequently, the single-particle spectrum exhibits an interaction-induced lifting of the spin-valley degeneracy, although some branches may still overlap or cross in the plotted band structure. The interaction-induced gap is 25.6\,meV. At filling factor $\nu=2$, the ground state is instead spin-degenerate but valley-polarized Chern insulator with $|C|=2$, with the characteristic order parameters $\left\langle s_z\right\rangle=\left\langle s_z\tau_z\right\rangle=0$ and $\left\langle\tau_z\right\rangle=2$. Here, the two spin species remain degenerate, while the valley degeneracy is lifted by exchange interactions. The interaction-induced gap is 32.8\,meV. In both cases, time-reversal symmetry is spontaneously broken, as evidenced by the finite Chern number and valley polarization.}

\begin{figure}[bth!]
%\begin{center}
    \includegraphics[width=5in]{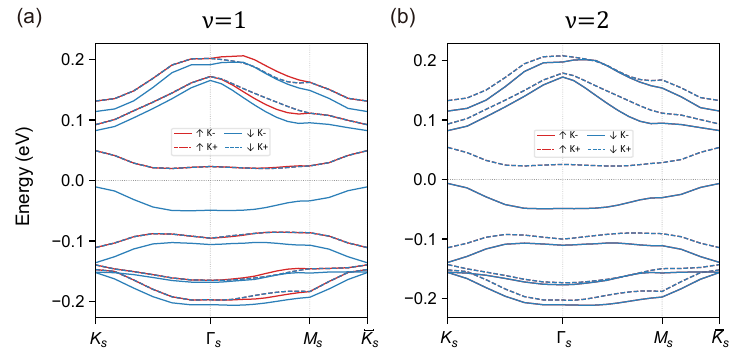}
\caption{Energy band structure for HF ground state of BLG-CrOCl heterostructure with $d_0=8$\,\AA at filling factor $\nu=1$ (a) with $E_\text{ext}=-0.1833$\,V/nm and $n_\text{tot}=1.6 \times 10^{12}$\, cm$^{-2}$,  at filling factor $\nu=2$ (b) with $E_\text{ext}=-0.1917$\,V/nm and $n_\text{tot}=2.25\times 10^{12}$\, cm$^{-2}$.}
\label{SI_7}
%\end{center}
\end{figure}

\BLUE{In \figurename ~\ref{SI_7}, we present the single-particle spectrum of Hartree-Fock ground states of BLG-CrOCl heterostructure with $d_0=8$\,\AA\,at integer filling factors $\nu=1$ (a) and $\nu=2$ (b). Similarly, these ground states are topologically correlated insulators stabilized by spontaneous symmetry breaking : At $\nu=1$, the system realizes a spin- and valley-polarized Chern insulator with Chern number $|C|=1$ and interaction-induced gap 30.0\,meV, while the ground state remains spin-degenerate but becomes valley-polarized at $\nu=2$, yielding a Chern insulator with $|C|=2$ and interaction-induced gap 29.0\,meV.}

\BLUE{For the HF ground states, the Chern number is computed using the self-consistently obtained HF Bloch wavefunctions. The integral is also performed over the superlattice mini Brillouin zone and calculated using the numerical method described above \cite{Chern-number-2005}, which does not require gauge continuity throughout the mini Brillouin zone.}

\BLUE{To rule out numerical artifacts as the origin of the nonzero Chern number, we have calculated the Berry-curvature distributions of the isolated Hartree-Fock bands shown in Fig. 2(a)-(b) of the main text, using the self-consistently obtained HF Bloch wavefunctions. The resulting distributions exhibit well‑resolved Berry curvature concentrated near the avoided crossings induced by the Wigner‑crystal superlattice, thus confirming the Chern‑insulating nature of the state.}

\begin{figure}[bth!]
%\begin{center}
    \includegraphics[width=5in]{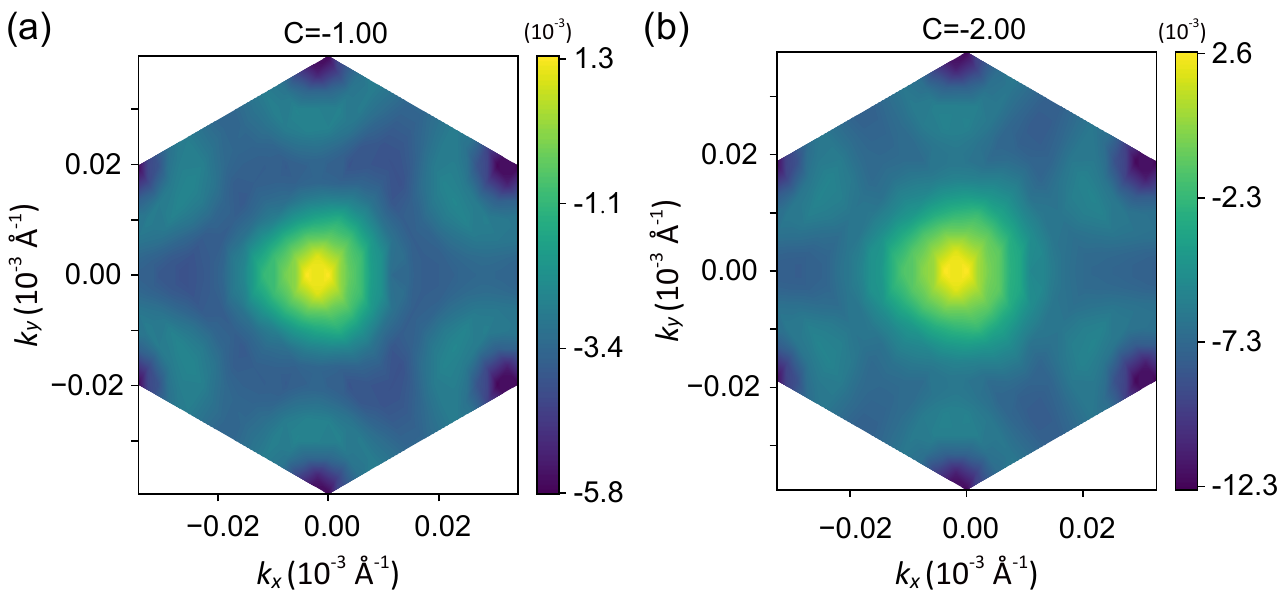}
\caption{The berry curvature distribution of the occupied bands in valley K for HF ground state of BLG: (a) at $E_\text{ext}=-0.3417 $\,$\text{V}/\text{nm}$ and $n_\text{tot}= 3.14\times10^{12}$\,cm$^{-2}$ with filling factor $\nu=1$; and (b) at $E_\text{ext}=-0.3542 $\,$\text{V}/\text{nm}$ and $n_\text{tot}= 3.78\times10^{12}$\,cm$^{-2}$ with filling factor $\nu=2$. }
\label{SI_8}
%\end{center}
\end{figure}

\begin{figure}[htb!]
    \centering
    \includegraphics[width=5in]{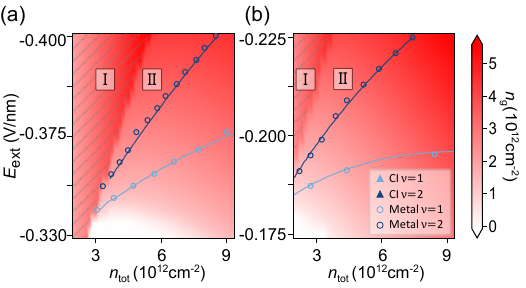}
    \caption{The HF ground states along iso-filling lines $\nu=1$ and $\nu=2$ are shown in (a) for BLG-CrOCl with $d_0=4.5$\,\AA, and in (b) for BLG-CrOCl with $d_0=8$\,\AA. The dielectric constant is set to $\epsilon_r=8$. The color coding represents carrier density $n_\text{g}$ of RMG, ``CI'' denotes Chern insulator state.} %Unshaded areas: charge transfer region with the LCB exhibiting a Chern number $|C|=1$ obtained from the non-interacting model with renormalized parameters. }
    \label{SI_9}
\end{figure}

\BLUE{ In the Fig.~2 of the main text and \figurename ~\ref{SI_5}-~\ref{SI_8}, the calculations is performed with a relative dielectric constant $\epsilon_r=4$.  And We have performed the same calculations for $\epsilon_r =8$, as \figurename ~\ref{SI_9} shown, where the Chern insulator state vanished for both $d_0=4.5$\,\AA\  and $d_0=8$\,\AA. }

\begin{figure}[htb!]
    \centering
    \includegraphics[width=5in]{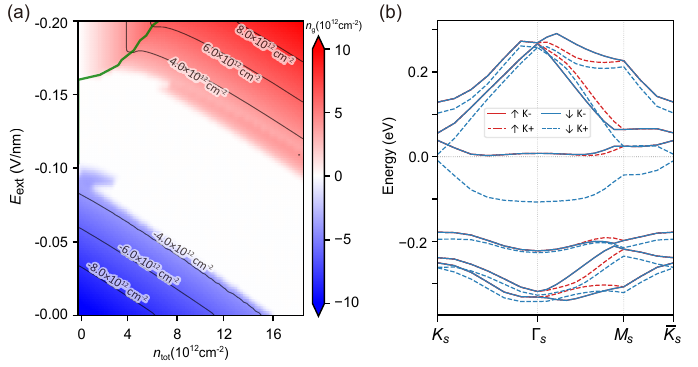}
    \caption{(a) Carrier densities of the TLG $n_{\text{g}}$ in theoretically modelled TLG-CrOCl heterostructures with interlayer distance $d_0=8.0$\,\AA. The black solid lines mark iso-doping levels in RMG. (b) HF single particle spectra for HF ground state of TLG at $E_\text{ext}=-0.1667 $\,$\text{V}/\text{nm}$ and $n_\text{tot}= 4.36\times10^{12}$\,cm$^{-2}$ with filling factor $\nu=1$. 
    } %Unshaded areas: charge transfer region with the LCB exhibiting a Chern number $|C|=1$ obtained from the non-interacting model with renormalized parameters. }
    \label{SI_10}
\end{figure}

\BLUE{To investigate the layer-number dependence, we generalize the above framework to the trilayer graphene(TLG)-CrOCl heterostructure. We first perform self-consistent electrostatic screening calculations with Fock corrections. The resulting phase diagram of the carrier density $n_\text{g}$ of TLG in the TLG-CrOCl heterostructure, as shown in \figurename ~\ref{SI_10}(a), reveals that the trilayer graphene shares the same qualitative physical behavior as the bilayer graphene. We then carry out band-projected self‑consistent Hartree-Fock calculations, which show that at filling factor $\nu=1$, the ground state of trilayer graphene spontaneously breaks time-reversal and stabilizes into a spin‑valley‑polarized Chern insulator as well.}

\section{\BLUE{Generalized susceptibility calculations}}
\label{sec: Generalized susceptibility calculations}
The Hartree-Fock approximation is biased toward symmetry-broken phases and may therefore overestimate the stability of the Chern insulator (CI) state. To overcome this limitation and provide an unbiased assessment, we perform random-phase approximation (RPA) calculations, which directly probe Fermi-surface quantum fluctuations and detect potential charge density wave (CDW) instabilities that could compete with the CI phase. In this section, we present the detailed RPA formalism used to evaluate the susceptibility and determine the leading instability modes.

Specifically, the bare susceptibility is defined in the original valley–spin–layer–sublattice basis of the continuum model as:
\begin{align}
    &\chi^0_{\mu\sigma l_\alpha\alpha \G, \mu'\sigma' l_\beta\beta\G+\Q;\mu\sigma l_\alpha'\alpha'\G',\mu'\sigma'l_\beta'\beta'\G'+\Q}(\btq,i\nu_n)\;\nn
    =&-\frac{1}{\beta}\sum_{i\omega_n}\int{\frac{dk^2}{(2\pi)^2} G^0_{\mu\sigma l_\alpha\alpha\G,\mu\sigma l_\alpha'\alpha'\G'}(i\omega_n, \btk)G^0_{\mu'\sigma' l_\beta'\beta'\G'+\Q,\mu'\sigma' l_\beta\beta\G+\Q}(i\omega_n+i\nu_n, \btk+\btq)},
\label{eq:chi0}
\end{align}
where $\mu$,$\mu'$ are the valley indices, $\sigma$,$\sigma'$ are the spin indices, $l_\alpha$,$l_\alpha'$,$l_\beta$,$l_\beta'$ are the layer indices, $\alpha$,$\alpha'$,$\beta$,$\beta'$ are the sublattice indices. $\G$,$\G'$ $\Q$ are the reciprocal lattice vectors of superlattice, $\btk$, $\btq$ are the wave vectors within the Brillouin zone of superlattice. $i\omega_n$, $i\nu_n$ are the Fermionic and Bosonic Matsubara frequencies, and $\beta=\frac{1}{k_BT}$, with $k_B$ denoting the Boltzmann constant and $T$ denoting the temperature. $G^0_{\mu\sigma l_\alpha\alpha\G,\mu\sigma l_\alpha'\alpha'\G'}(i\omega_n, \btk)$ denotes the non-interacting single-particle Green’s function expressed
in the original basis of the continuum model:
\begin{equation}
    G^0_{\mu\sigma l_\alpha\alpha\G,\mu\sigma l_\alpha'\alpha'\G'}(i\omega_n, \btk) = \sum_{n\in \text{flat}}{\frac{C^*_{\mu l_\alpha\alpha\G,n\btk}C_{\mu l_\alpha'\alpha'\G', n\btk}}{i\omega_n-E_{\mu,n\btk}}}
    \label{eq:green}
\end{equation}
where the $C_{\mu l_\alpha\alpha\mathbf{G},n\kt}$ denotes the non-interacting wavefunction at $\btk$, and $n$ is the band index. Since the Coulomb interaction effects are most prominent for the flat bands and we are interested in the instability modes driven by quantum fluctuations of the flat bands, the summation of the band index $n$ in Eq.~(\ref{eq:green}) is restricted   to the flat-band subspace. $E_{\mu,n\btk}$ denotes the flat-band dispersion \textit{including the Coulomb potentials from the occupied remote energy bands}. It is worthwhile to note that, the Coulomb potentials from the remote energy bands make the flat bands much more dispersive, which somehow justifies the RPA treatment in calculating the generalized susceptibility tensor.
Plugging Eq.~(\ref{eq:green}) into Eq.~(\ref{eq:chi0}), and carrying out the summation over Matsubara frequency, one obtains
\begin{align}
&\chi^{0}_{\mu\sigma l_\alpha\alpha\mathbf{G},\mu'\sigma'l_\beta\beta\;\mathbf{G}+\mathbf{Q}\;;\;\mu\sigma l_\alpha'\alpha'\mathbf{G}',\mu'\sigma'l_\beta'\beta'\;\mathbf{G}'+\mathbf{Q}}(\widetilde{\mathbf{q}},i\nu_n)\;\nn
=&\int\frac{dk^2}{(2\pi)^2}\sum_{n,m}\,C^{*}_{\mu l_\alpha\alpha\mathbf{G},n\kt}\,C_{\mu l_\alpha'\alpha'\mathbf{G}',n\kt}\,C^{*}_{\mu'l_\beta'\beta'\;\mathbf{G}'+\mathbf{Q},m\kt+\qt}\,C_{\mu'l_\beta\beta\;\mathbf{G}+\mathbf{Q},m\kt+\qt}\,\frac{f(E_{\mu,n\widetilde{\k}})-f(E_{\mu',m\widetilde{\k}+\widetilde{\mathbf{q}}})}{E_{\mu',m\widetilde{\k}+\widetilde{\mathbf{q}}}-E_{\mu,n\widetilde{\k}}-i\nu_n}
\end{align}
Such a susceptibility tensor defined in the original basis characterizes the intrinsic Fermi-surface fluctuations in the valley-spin-layer-sublattice space. In order to describe the possible spontaneous symmetry-breaking states, here we only consider the zero-frequency susceptibility with $i\nu_n\!=\!0$, and keep the full moir\'e wavevector dependence: 
\begin{equation}
\chi^{0}_{\mu\sigma l_\alpha\alpha\mathbf{G},\mu'\sigma'l_\beta\beta\;\mathbf{G}\;;\;\mu\sigma l_\alpha'\alpha'\mathbf{G}',\mu'\sigma'l_\beta'\beta'\;\mathbf{G}'}(\btq,\mathbf{Q})\equiv\chi^{0}_{\mu\sigma l_\alpha\alpha\mathbf{G},\mu'\sigma'l_\beta\beta\;\mathbf{G}+\mathbf{Q}\;;\;\mu\sigma l_\alpha'\alpha'\mathbf{G}',\mu'\sigma'l_\beta'\beta'\;\mathbf{G}'+\mathbf{Q}}(\btq,i\nu_n\!=\!0)\;.
\end{equation}
Electron-electron Coulomb interactions may greatly enhance the susceptibility tensor and drive a second-order phase transition via spontaneous symmetry breaking in the valley-spin-layer-sublattice space. In particular, we consider the dominant intravalley Coulomb interactions as given by Eq.~(\ref{eq:h-intra}). The two-particle Coulomb scattering processes involve  those from both direct  and  exchange Coulomb interactions, which can be written in  matrix form as:
\begin{align}
&\mathbb{U}(\btq,\Q)_{\mu_{\alpha}\sigma_{\alpha}l_\alpha\alpha\mathbf{G},\mu_{\beta}\sigma_{\beta}l_\beta\beta\mathbf{G}\;;\;\mu_{\alpha}'\sigma_{\alpha}'l_\alpha'\alpha'\mathbf{G}',\mu_{\beta}'\sigma_{\beta}'l_\beta'\beta'\mathbf{G}'}\;\nn
=&V(\vert\btq+\Q\vert)\,\delta_{\mu_{\alpha}\mu_{\beta}}\delta_{\sigma_{\alpha}\sigma_{\beta}}\delta_{l_\alpha l_\beta}\delta_{\alpha\beta}\delta_{\mu_{\alpha}'\mu_{\beta}'}\delta_{\sigma_{\alpha}'\sigma_{\beta}'}\delta_{l_\alpha' l_\beta'}\delta_{\alpha'\beta'}-V(\vert\btk+\G-\btk'-\G'\vert)\,\delta_{\mu_{\alpha}\mu_{\alpha}'}\delta_{\mu_{\beta}\mu_{\beta}'}\delta_{\sigma_{\alpha}\sigma_{\alpha}'}\delta_{\sigma_{\beta}\sigma_{\beta}'}\delta_{l_\alpha l_\alpha'}\delta_{l_\beta l_\beta'}\delta_{\alpha\alpha'}\delta_{\beta\beta'}\;\nn
\approx &V(\vert\widetilde{\mathbf{q}}+\mathbf{Q}\vert)\,\delta_{\mu_{\alpha}\mu_{\beta}}\delta_{\sigma_{\alpha}\sigma_{\beta}}\delta_{l_\alpha l_\beta}\delta_{\alpha\beta}\delta_{\mu_{\alpha}'\mu_{\beta}'}\delta_{\sigma_{\alpha}'\sigma_{\beta}'}\delta_{l_\alpha' l_\beta'}\delta_{\alpha'\beta'}-V(\vert\mathbf{G}-\mathbf{G}'\vert)\,\delta_{\mu_{\alpha}\mu_{\alpha}'}\delta_{\mu_{\beta}\mu_{\beta}'}\delta_{\sigma_{\alpha}\sigma_{\alpha}'}\delta_{\sigma_{\beta}\sigma_{\beta}'}\delta_{l_\alpha l_\alpha'}\delta_{l_\beta l_\beta'}\delta_{\alpha\alpha'}\delta_{\beta\beta'}
\label{eq:coulomb-matrix}
\end{align}
Again, $\{\mu_{\alpha},\mu_{\beta},\mu_{\alpha'},\mu_{\beta '}\}$, $\{\sigma_{\alpha},\sigma_{\beta},\sigma_{\alpha'},\sigma_{\beta '}\}$, $\{l_\alpha,l_\beta,l_\alpha',l_\beta'\}$ and $\{\alpha,\beta,\alpha',\beta'\}$ denote the valley, spin, layer and sublattice indices respectively, and $V(\q)$ is the Coulomb interaction described as Eq.~(\ref{eq:Vq}). It is important to note that, in the last line of Eq.~(\ref{eq:coulomb-matrix}), we have made an approximation that the amplitude of the exchange Coulomb interaction is only dependent on the transfer of reciprocal vector  $\G-\G'$, neglecting the transfer of the wavevector $\btk-\btk'$ within the mini Brillouin zone of superlattice. This is an excellent approximation given the small size of the mini Brillouin zone of superlattice with lattice constant $L_s \ge 100$\,\AA. Then we calculate the interaction-renormalized generalized susceptibility tensor with random phase approximation. By virtue of the approximation made in Eq.~(\ref{eq:coulomb-matrix}), the RPA susceptibility can be written in a succinct matrix form:
\begin{equation}
\hat{\chi}^{\rm{RPA}}(\widetilde{\mathbf{q}},\mathbf{Q})=\hat{\chi}^{0}(\qt,\mathbf{Q})\,\cdot\,(1+\mathbb{U}(\qt,\mathbf{Q})\cdot\hat{\chi}^{0}(\widetilde{q},\mathbf{Q}))^{-1}
\end{equation}
where $\hat{\chi}^{0}(\widetilde{q},\mathbf{Q}))$ is the matrix of bare susceptibility, whose matrix element is given in Eq.~(\ref{eq:chi0}), and $\hat{\chi}^{\rm{RPA}}(\widetilde{\mathbf{q}},\mathbf{Q})$ is the RPA susceptibility tensor defined in the same basis as  $\hat{\chi}^{0}(\qt,\mathbf{Q}))$ and $\mathbb{U}(\widetilde{\mathbf{q}},\mathbf{Q})$. In the end, we sum over all the transferred reciprocal vectors $\mathbf{Q}$, and define the RPA susceptibility at a moir\'e wavevector $\widetilde{\mathbf{q}}$ (within moir\'e Brillouin zone) as:
\begin{equation}
\hat{\chi}^{\rm{RPA}}(\widetilde{\mathbf{q}})=\sum_{\mathbf{Q}}\,\hat{\chi}^{\rm{RPA}}(\widetilde{\mathbf{q}},\mathbf{Q})
\label{eq:chiRPA}
\end{equation}
which captures the Fermi-surface quantum fluctuations in the valley-spin-layer-sublattice space contributed by the flat bands.  One can diagonalize $\hat{\chi}^{\rm{RPA}}(\widetilde{\mathbf{q}})$ at each moir\'e wavevector $\widetilde{\mathbf{q}}$, and any instability modes with diverging eigenvalues would indicate the tendency of forming a symmetry-breaking density-wave state with wavevector $\widetilde{\mathbf{q}}$. The  eigenvectors of the diverging modes would correspond to the order parameters of the possible density-wave states.

\begin{figure}[htb!]
    \centering
    \includegraphics[width=5in]{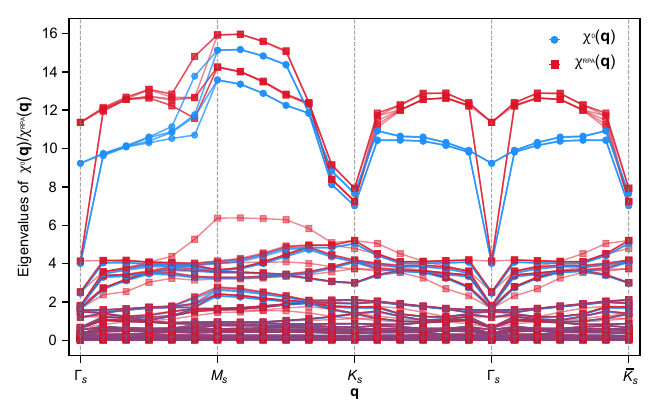}
    \caption{The bare susceptibility (blue) and the RPA susceptibility (red) at $\nu=1$ filling with $E_\text{ext}=-0.3414 $\,$\text{V}/\text{nm}$ and $n_\text{tot}= 3.14\times10^{12}$\,cm$^{-2}$. }
    \label{SI_11}
\end{figure}

We present the eigenvalues of the bare susceptibilities and the interaction-renormalized RPA susceptibilities in the \figurename ~\ref{SI_11}.  we do not find prominent instability modes along the high-symmetry path, indicating that the Chern insulator state remains stable against potential symmetry-breaking fluctuations within the considered system.

\section{Full-band Hartree-Fock Method}
\label{sec: Full-band Hartree-Fock Method}
We now go beyond the ``Born-Oppenheimer-type'' treatment and turn to the full-band Hartree-Fock (HF) method, which treats electrons transferred to the substrate and within the RMG on an equal footing. Technically, the full-band HF method is unavoidable when treating band dispersions with substantially different energy scales, as in our case. In particular, no physically sensible band cutoff exists for a band-projected HF calculation, which again necessitates a full-band HF treatment.

Low-energy electrons within RMG are described by the standard non-interacting $\bk \cdot \p$ model, while the electrons transferred to the substrate are modeled as a two-dimensional electron gas with a parabolic band dispersion. Owing to the large interlayer separation $d_0$ and lattice mismatch, direct interlayer electron hopping is negligible; however, long-range Coulomb interactions between the two subsystems remain significant. We introduce a layer-dependent Coulomb potential to distinguish the interlayer interactions and intralayer interactions. The total Hamiltonian of the coupled RMG-insulator substrate heterostructure is expressed as:
\begin{subequations}
    \begin{align}
        \hat{H}^0 &=  \sum_{\lambda\lambda',\sigma\bk} \epsilon_{\lambda,\lambda'}^{\sigma}(\bk) \hat{c}^{\dagger}_{\lambda\sigma}(\bk)\hat{c}_{\lambda'\sigma}(\bk), \label{Za} \\ 
        \hat{H}_{\text{int}} &= \frac{1}{2N_s}\sum_{\substack{\sigma\sigma' \\ \lambda\lambda'}}\sum_{\bk\bk'\q} {V_{\lambda, \lambda'}(\q)} \notag \times \hat{c}^\dagger_{\lambda\sigma}(\bk+\q)\hat{c}^\dagger_{\lambda'\sigma'}(\bk'-\q)\hat{c}_{\lambda'\sigma'}(\bk')\hat{c}_{\lambda\sigma}(\bk), \label{Zb}
    \end{align}
\end{subequations}
where the composite index $\lambda \equiv (\mu, l, \alpha)$ collectively labels the valley, layer, and sublattice degrees of freedom, and $\sigma$ denotes the spin index. We consider the 2D electron gas on the substrate' side with a locked valley-layer-sublattice degree of freedom. Accordingly, $\lambda = -1$ is assigned to the surface the of the substrate, and non-negative values $\lambda \geq 0$ correspond to graphene layers of the RMG. Taking the BLG-substrate heterostructure with two-sublattice and two-valley degrees of freedom as an example, the relation $\lambda = 4\mu + 2l + \alpha$ establishes the correspondence between an electron with index $\lambda$ and an electron residing in layer $l$, valley $\mu$, and sublattice $\alpha$. The operator $\hat{c}^\dagger_{\lambda\sigma}(\bk)$ $(\hat{c}_{\lambda\sigma}(\bk))$ creates (annihilates) an electron with wave vector $\bk$ expanded around the CBM of the substrate or the charge neutrality point of RMG, carrying the composite index $\lambda$ and spin $\sigma$. Again taking the BLG-based charge transfer heterostructure as an example, the expression of the non-interacting kinetic term $\epsilon_{\lambda\lambda', \sigma\bk}$ is :
\begin{align}    
    \epsilon_{\sigma\bk} = 
    \begin{pmatrix}
        \frac{\hbar^2 k^2}{2m^*}+E_{CBM} & 0 & 0 & 0 & 0 \\
        0 & h^{0, +}_{\text{intra}} & (h^{0, +}_{\text{inter}})^\dagger & 0 & 0 \\
        0 & h^{0, +}_{\text{inter}} & h^{0, +}_{\text{intra}} & 0 & 0 \\
        0 & 0 & 0 & h^{0, -}_{\text{intra}} & (h^{0, -}_{\text{inter}})^\dagger \\
        0 & 0 & 0 & h^{0, -}_{\text{inter}} & h^{0, -}_{\text{intra}}    
    \end{pmatrix}\;.
\end{align}
The effective mass $m^*$ and $E_{\text{CBM}}$, the initial energy of the CBM of substrate relative to the charge neutrality point of RMG,  are set to be tunable parameters in this section. The interaction Hamiltonian $\hat{H}_{\text{int}}$ not only contains the e-e Coulomb interaction between electrons on graphene side, but also interactions between electrons in the substrate and coupling between them. 
As discussed previously, we adopted a layer-dependent screened Coulomb potential $V_{ll'}(\bq)$ Eq.~(\ref{eq:Vq}) to distinguish the difference between intralayer Coulomb interaction and interlayer Coulomb interaction. In our numerical calculations, We begin with trial initial states in which translational symmetry is weakly broken by introducing a trial superlattice potential, whose lattice constant is determined by $\sqrt{3}L_s^2n_\mathrm{sub}/2=2$, and then perform self-consistent Hartree-Fock iterations until convergence is reached. $N_s$ is the number of the unit cell in the system. Hence, wave vectors are folded into the corresponding mini Brillouin zone by $\bk = \tilde{\bk}+\G$  and $\q = \tilde{\q}+\Q$. $\G$ and $\Q$ are reciprocal lattice vectors of mini Brillouin zone. Due to exponentially decaying potential, we only consider the dominant intravalley interaction. Applying Hartree-Fock approximation to the full band interaction Hamiltonian, we get:
\begin{figure*}[!htb]
    \centering
    \begin{equation}
        \hat{V}_H^{full}=\frac{1}{N_s}\sum_{\substack{\sigma\sigma'\\ \lambda\lambda'}}\sum_{\tilde{\bk}\tilde{\bk}'}\sum_{\G,\G',\Q} V_{\lambda\lambda'}(\Q)
        \left \langle \hat{c}^\dagger_{\sigma'\lambda'\G'-\Q, \tilde{\bk}'}\hat{c}_{\sigma'\lambda'\G', \tilde{\bk}'} 
        \right \rangle \hat{c}^\dagger_{\sigma\lambda\G+\Q, \tilde{\bk}} \hat{c}_{\sigma\lambda\G, \tilde{\bk}}
    \end{equation}

    \begin{equation}
        \begin{aligned}
            \hat{V}_F^{full}=-\frac{1}{N_s}\sum_{\substack{\sigma\sigma'\\ \lambda\lambda'}}\sum_{\tilde{\bk}\tilde{\bk}'}\sum_{\substack{\G,\G',\Q}} V_{\lambda\lambda'}(\tilde{\bk}'-\tilde{\bk}+\G'-\G+\Q)  \times \left \langle \hat{c}^\dagger_{\sigma\lambda\G'+\Q, \tilde{\bk}'}\hat{c}_{\sigma'\lambda'\G', \tilde{\bk}'} \right \rangle \hat{c}^\dagger_{\sigma'\lambda'\G-\Q, \tilde{\bk}} \hat{c}_{\sigma\lambda\G, \tilde{\bk}}
        \end{aligned}
    \end{equation}
\end{figure*}

\clearpage
%\endcontents
\bibliography{reference}

\end{document}